\documentclass[12pt,letterpaper]{article}
\usepackage{amsmath,amssymb,array,calc,rotating,epsfig,psfrag,amscd, cite}

\makeatletter
\renewcommand\section{\@startsection {section}{1}{\z@}%
                                   {-3.5ex \@plus -1ex \@minus -.2ex}%nn
                                   {2.3ex \@plus.2ex}%
                                   {\normalfont\large\bfseries}}
\renewcommand\subsection{\@startsection{subsection}{2}{\z@}%
                                     {-3.25ex\@plus -1ex \@minus -.2ex}%
                                     {1.5ex \@plus .2ex}%
                                     {\normalfont\bfseries}}
\makeatother

\let\non\nonumber

\let\s=\sigma

\newcommand{\bea}{\begin{eqnarray}}
\newcommand{\eea}{\end{eqnarray}}
\newcommand{\be}{\begin{equation}}
\newcommand{\ee}{\end{equation}}

\newcommand{\Z}{{\mathbb Z}}

\newcommand{\p}{\partial}

\newcommand{\C}[1]{$(\ref{#1})$}

\def\IZ{\relax\ifmmode\mathchoice
{\hbox{\cmss Z\kern-.4em Z}}{\hbox{\cmss Z\kern-.4em Z}}
{\lower.9pt\hbox{\cmsss Z\kern-.4em Z}} {\lower1.2pt\hbox{\cmsss
Z\kern-.4em Z}}\else{\cmss Z\kern-.4em Z}\fi}
\def\IR{\relax{\rm I\kern-.18em R}}

\def\one{{\hbox{ 1\kern-.8mm l}}}

\newlength{\bredde}
\def\slash#1{\settowidth{\bredde}{$#1$}\ifmmode\,\raisebox{.15ex}{/}
\hspace*{-\bredde} #1\else$\,\raisebox{.15ex}{/}\hspace*{-\bredde}
#1$\fi}

\newsavebox{\zzzbar}
\sbox{\zzzbar}
  {\setlength{\unitlength}{0.9em}
  \begin{picture}(0.6,0.7)
  \thinlines
  \put(0,0){\line(1,0){0.6}}
  \put(0,0.75){\line(1,0){0.575}}
  \multiput(0,0)(0.0125,0.025){30}{\rule{0.3pt}{0.3pt}}
  \multiput(0.2,0)(0.0125,0.025){30}{\rule{0.3pt}{0.3pt}}
  \put(0,0.75){\line(0,-1){0.15}}
  \put(0.015,0.75){\line(0,-1){0.1}}
  \put(0.03,0.75){\line(0,-1){0.075}}
  \put(0.045,0.75){\line(0,-1){0.05}}
  \put(0.05,0.75){\line(0,-1){0.025}}
  \put(0.6,0){\line(0,1){0.15}}
  \put(0.585,0){\line(0,1){0.1}}
  \put(0.57,0){\line(0,1){0.075}}
  \put(0.555,0){\line(0,1){0.05}}
  \put(0.55,0){\line(0,1){0.025}}
  \end{picture}}

\newcommand{\ena}{\end{eqnarray}}
\newcommand{\beqa}{\begin{eqnarray}}
\newcommand{\eeqa}{\end{eqnarray}}

\newcommand{\g}{\gamma}

\def\g{\gamma}

\def\s{\sigma}

\begin{document}
\begin{titlepage}

\begin{center}

%{September 26, 2005}
\today
\hfill         \phantom{xxx}

\hfill EFI-08-20 

\vskip 2 cm
{\Large \bf Recursion Relations from Space-time Supersymmetry}\\
\vskip 1.25 cm { Anirban Basu\footnote{email address: abasu@ias.edu}$^{a}$ and Savdeep Sethi\footnote{email address:
 sethi@theory.uchicago.edu}}$^{b}$\\
{\vskip 0.5cm $^{a}$ Institute for Advanced Study, Princeton, NJ 08540, USA\\}

{\vskip 0.5cm $^{b}$ Enrico Fermi Institute, University of Chicago,
Chicago, IL
60637, USA\\}

\end{center}
\vskip 2 cm

\begin{abstract}
\baselineskip=18pt

We describe a method for obtaining relations between higher 
derivative interactions in supersymmetric effective actions. The 
method extends to all orders in the momentum expansion. As an 
application, we consider the string coupling dependence of the $\hat{G}^{2k} 
\lambda^{16}$ interaction in type IIB string theory. 
Using supersymmetry, we show that each of these 
interactions satisfies a Poisson equation on the moduli space with sources 
determined by lower momentum interactions. We argue that these 
protected couplings are only renormalized by a finite number of 
string loops together with non-perturbative terms. Finally, we 
explore some consequences of the Poisson equation for low values of 
$k$.

\end{abstract}

\end{titlepage}

\pagestyle{plain}
%\baselineskip=18pt
% Try a wider skip
\baselineskip=19pt
%%%%%%%%%%%%%%%%%%%%%%%%%%%%%%%%%%%%%%%%%%%%%%%%%%%%%%%%%%%%%%%%%%%%%%%%%%%%%%
\section{Introduction}

The string effective action has an intricate and beautiful 
structure. The higher derivative interactions in the effective 
action play an important role in understanding the ultraviolet 
structure of the theory beyond both the supergravity approximation 
and the perturbative string approximation. These interactions also 
play a role in resolving singularities of classical supergravity 
solutions and in improving our understanding of non-perturbative 
dualities. 

A direct study of the space-time action is, in many ways, 
complimentary to S-matrix computations in perturbative string 
theory. The former gives non-perturbative results in the string 
coupling but usually to a fixed order in the momentum expansion. The 
latter gives results to all orders in the momentum expansion but to 
a fixed order in the string coupling expansion. Combining the data 
from both approaches will help determine the structure of the 
complete non-perturbative S-matrix.

In this work, we will be concerned with the 1PI effective action 
which is duality invariant. Our main result will be to explain how 
to derive recursion relations relating special higher momentum 
interactions. The method applies quite generally though we will 
focus on the case of type IIB string theory in ten dimensions. 

We will show that each special interaction satisfies a Poisson 
equation on the moduli space with sources at most cubic in 
the couplings of lower momentum 
operators. For other choices of couplings, this structure will 
generalize to a system of equations with sources. We will focus on 
the simplest examples in this work which satisfy second order 
equations. As a consequence of this constraint, these interactions 
do not receive string loop contributions beyond a certain 
loop order extending the result of~\cite{Berkovits:2006vc}. They do, 
however, receive non-perturbative corrections which might be 
interpretable as coming from D-instantons in some cases, 
or bound-states of D-instantons and D-anti-instantons. This leads to a quite 
beautiful interplay between modular forms and space-time couplings. 
It would be exciting to relate the non-perturbative effects to 
twisted partition functions of brane systems along the lines 
of~\cite{Green:1997tv}.

There has been considerable work devoted to understanding higher derivative interactions in theories with maximal supersymmetry in different dimensions. Most of the analysis involves the first few terms in 
the $\alpha'$ expansion of the effective action; for a selection of papers, see~\cite{Green:1997tv, Green:1997di, Green:1997as,Kiritsis:1997em,Green:1997me,
Pioline:1998mn,Green:1998by,Green:1999pu,Obers:1999es,Sinha:2002zr,Berkovits:2004px,DHoker:2005jc,DHoker:2005ht,Green:2005ba,Berkovits:2006vc,Green:2006gt,Basu:2007ru,Basu:2007ck}. Of particular interest to us are interactions in ten-dimensional type IIB string theory of the form
\be f_k^{(0,0)} (\tau, \bar\tau) D^{2k} {\mathcal{R}}^4 \ee
where $f_k^{(0,0)} (\tau, \bar\tau)$ is a string coupling-dependent coefficient function. For low values of $k$, a great deal is known or conjectured about these $f_k^{(0,0)}$~\cite{Green:1997as,Green:1998by, Green:1999pu, Sinha:2002zr, Green:2005ba}, 
\be \label{knownstuff} f_0^{(0,0)} (\tau,\bar\tau) = E_{3/2} (\tau,\bar\tau) ,  \quad
f_2^{(0,0)}(\tau,\bar\tau)= E_{5/2} (\tau,\bar\tau), \ee
while $f_3^{(0,0)}$ satisfies
\be 4 \tau_2^2 \frac{\p^2}{\p \tau \p \bar\tau} f_3^{(0,0)} (\tau,\bar\tau)= 12 f_3^{(0,0)} (\tau,\bar\tau) - 6 
\Big( f_0^{(0,0)} (\tau,\bar\tau) \Big)^2. \ee
Here $E_s (\tau,\bar\tau)$ is the non--holomorphic Eisenstein series given in Appendix~\ref{summarymodular}.

\subsection{A sketch of the argument}
\label{sketch}

Before we delve into a complete analysis, it worth sketching schematically the basic
idea about why there should recursion relations relating an infinite set of 
higher momentum operators. Let us first recall that type IIB supergravity
enjoys a $U(1)$ symmetry which is broken in string theory by non-perturbative
interactions like those mediated by D-instantons. For a review, see~\cite{Green:1999qt}.
Let us recall the $U(1)$ charge assignment to the various fields and parameters of type IIB 
supergravity given by~\cite{Schwarz:1983qr,Howe:1983sra}, 
\be
[G]=1, \quad [\lambda] = {3\over 2}, \quad [F_5]=0,\quad [g_{\mu\nu}] =0, 
\quad[\psi]={1\over 2},\quad [ \partial \tau ] = 2, \quad [\epsilon]={1\over 2},
\ee
where $\epsilon$ is the supersymmetry transformation parameter, $\lambda$ is the dilatino, $\psi$  the gravitino, $g_{\mu\nu}$ the metric, $\tau$ the string coupling, $F_5$ the self-dual $5$-form field strength and $G$ the complex $3$-form field strength. A field carrying $U(1)$ charge $q$ 
has modular weight $(-q/2,q/2)$ under $SL(2,\mathbb{Z})$ transformations. Some relevant properties
of modular forms are summarized in Appendix~\ref{summarymodular}. 

Now at the eight derivative level, there is a nice superfield formalism that relates couplings like
\be\label{chains}
f^{(0,0)}_0 (\tau, \bar\tau) {\mathcal R}^4 + \ldots + f^{(12,-12)}(\tau, \bar\tau)  \lambda^{16}. 
\ee
Supersymmetry naturally constrains the coupling containing the most fermions which determines 
$f^{(12,-12)}$ as shown in~\cite{Green:1998by}\ extending the arguments of~\cite{Paban:1998ea, Paban:1998qy, Sethi:1999qv}. The $f^{(12,-12)}$ coefficient function for $\lambda^{16}$ is proportional to 
\be
D_{11} \cdots D_0 f^{(0,0)}_0 (\tau,\bar\tau) 
\ee
where $f^{(0,0)}_0(\tau, \bar\tau)$ appears in~\C{knownstuff}\ and $D_m$ are modular covariant derivatives defined in Appendix~\ref{summarymodular}. 

This line of reasoning leads to constraints so long as you have moduli-dependent fermionic couplings which do not vanish when you apply supersymmetry to the moduli-dependent coefficients. However, at some point in the momentum expansion, we will simply run out of fermions 
to use to build interactions so we expect this kind of argument to extend to a finite (but high) order in the momentum expansion. 

Now this is a little too simplistic. A complete analysis of the supersymmetry constraints for maximally supersymmetric $0+1$-dimensional Yang-Mills was performed in~\cite{Sethi:2004np}. In that analysis, relations were found to all orders in the momentum expansion but they related a special coupling at order $2k$ in the derivative expansion to couplings (both protected and unprotected) at order $2(k-1)$. What we will show in this work is that there is a much richer and more powerful set of recursion relations when one considers field theory rather than quantum mechanics.

We are going to consider operators of the form $\hat{G}^{2k} \lambda^{16}$ in type IIB string theory. The case $\hat{G}^{4} \lambda^{16}$ was already studied in~\cite{Sinha:2002zr}\ where the coefficient function was argued to be proportional to
\be
D_{13} \cdots D_0 f_2^{(0,0)}(\tau,\bar\tau)
\ee
where $f_2^{(0,0)}(\tau,\bar\tau)$ is given in~\C{knownstuff}.

 We expect these couplings to be related to $D^{2k} {\mathcal R}^4$ by supersymmetry giving a schematic structure
\be \label{schematicstructure}
f^{(0,0)}_k (\tau, \bar\tau)D^{2k}  {\mathcal R}^4 + \ldots + f^{(12+k,-12-k)} (\tau, \bar\tau)\hat{G}^{2k}\, \lambda^{16}
\ee
analogous to~\C{chains}. Unfortunately, the relation between these interactions cannot be obtained from any (simple) superspace argument and it 
remains an outstanding question to obtain a precise relation. 

To see what is special about these particular interactions, note that  the supercovariant combination, $\hat{G}_{\mu\nu\rho}$, contains a chiral gravitino coupling $\bar\psi_{[\mu}^* \g_\nu \psi_{\rho]}$. We can expand powers of $\hat{G}$ as follows:
\be 
\hat{G}^{2k} = (\psi \psi)^{2N} G^{2(k-N)} + \ldots. 
\ee 
Here $2N$ denotes the largest non-vanishing power of $(\psi \psi)$ which we will determine later. The exact value of $N$ is unimportant. The omitted terms involve less fermions.  Expanding the chiral space-time couplings of interest gives
\be \label{chiral}
 f^{(12+k,-12-k)} (\tau, \bar\tau)\lambda^{16}\, \hat{G}^{2k} =  f^{(12+k,-12-k)} (\tau, \bar\tau) \lambda^{16} (\psi \psi)^{2N} G^{2(k-N)} + \ldots.
\ee 
Again the omitted terms have fewer fermions. Now the key point is what happens under a variation of $\bar\tau$ which
gives,
\bea
\delta \left(  f^{(12+k,-12-k)} (\tau, \bar\tau) \lambda^{16} (\psi \psi)^{2N} G^{2(k-N)} \right) &=& \bar\partial  f^{(12+k,-12-k)} (\tau, \bar\tau)  \epsilon^* \lambda^* \times \cr && \lambda^{16} (\psi \psi)^{2N} G^{2(k-N)} + \ldots.
\eea
To obtain a constraint, this term cannot mix with any higher fermion term of the same order in the momentum expansion. Such a term would have the schematic structure
\be
\lambda^* \lambda^{16} (\psi \psi)^{2N} G^{2(k-N)-1} \times F. 
\ee
The fermion $F$ must vary into $G$. After quickly perusing Appendix~\ref{susysummary}, we see that the only fermions with this property are $(\lambda, \psi)$ but the resulting coupling then vanishes by Fermi statistics. Therefore the coupling~\C{chiral}\ should be special and constrained.

Now in this argument we have ignored several issues: the first is mixing with other couplings with the same number of fermions; the second is mixing with terms in the supergravity action via higher derivative corrections to the supersymmetry transformations; the final issue is mixing with source terms from lower derivative interactions (but still beyond supergravity) again via corrections to the  supersymmetry transformations. We will address all of these issues in the bulk of this work but the above argument gives the core reason to expect constraints. It is very general. We expect similar reasoning to apply to protected couplings in  theories with N=4, N=2 and perhaps even N=1 supersymmetry. 

\subsection{A brief summary}

Let us summarize the results. In addition to the $\hat{G}^{2k} \lambda^{16}$ interaction with coefficient function $f^{(12 +k, -12-k)}(\tau,\bar\tau)$, we also need to consider $\hat{G}^{2k} \lambda^{15} \g^\mu \psi_\mu^*$ and $\hat{G}^{2(k-1)} (\hat{G}^{\mu\nu\rho} \hat{G}_{\mu\nu\rho}^*) \lambda^{16}$ which have the same coefficient function $f^{(11+ k',-11-k')}(\tau,\bar\tau)$. These modular forms satisfy two coupled equations which are derived in section~\ref{derivation}. The first equation is a Poisson equation  on the fundamental domain of $SL(2,\mathbb{Z})$ for $f^{(12 +k, -12-k)}(\tau,\bar\tau)$,
\bea \label{resultone}
&&D_{11 +k} \bar{D}_{-(12 + k)} f^{(12 +k, -12-k)} = a_k f^{(12+k, -12-k)}  \non \\
&&+D_{11 +k} \sum_{k'} 
\Big( b_{kk'} f^{(11+ k',-11-k')} f^{(k-k',k'-k)} + c_{kk'} f^{(12+ k',-12-k')} f^{(k-k'-1,k'-k+1)} \Big),\qquad \eea
with eigenvalue $a_k$ and source terms with coefficients $(b_{kk'}, c_{kk'})$. The source terms arise from interactions in the effective action beyond supergravity but at an order $k'< k$ in the effective action. The source terms themselves are special and only involve interactions related to $D^{2k'} \mathcal{R}^4$ by supersymmetry. 

The second equation has a similar form, 
\bea \label{resulttwo}
&&\bar{D}_{-(12+k)} D_{11 +k} f^{(11+k, -11-k)} =  a_k f^{(11+k , -11-k)} \non \\
&&+ \sum_{k'} \Big( d_{kk'} f^{(11+ k', -11 -k')} f^{(k-k', k'-k)} +
e_{kk'} f^{(12+ k',-12-k')} f^{(k-k'-1,k'-k+1)}\Big), \quad\eea
with sources appearing with coefficients $(d_{kk'}, e_{kk'})$. In principle, all the numerical coefficients $(a_k, b_{kk'}, c_{kk'}, d_{kk'}, e_{kk'})$ are determined by supersymmetry.  In practice, it is simpler to fix their values for specific choices of $k$ by using additional data from perturbative string and supergravity computations.  It is possible that some underlying topological string theory might be useful for determining the coefficients. Though the equations have an intricate recursive structure, they involve a very specific pattern of interactions that is highly constrained. Also note that there can be different equations for each possible space-time structure appearing in $\hat{G}^{2k}$. However, they all have the form of~\C{resultone}\ and~\C{resulttwo}, just with different coefficients.

The existence of the recursion relations~\C{resultone}\ and~\C{resulttwo}\ leads to a variety of results for these protected couplings which are explored in section~\ref{implications}. The protected couplings receive only a finite number of perturbative string loop contributions regardless of how large $k$ might be. We expect a version of this result to also hold for $D^{2k} {\mathcal R}^4$ and many other special couplings using supersymmetry to relate the couplings with the schematic structure depicted in~\C{schematicstructure}. There should be a sort of supermultiplet of couplings built from $\hat{G}^{2k} \lambda^{16}$ which enjoys special renormalization properties. 

Using supersymmetry to chain from the maximal fermion interactions to couplings with fewer or no fermions like $D^{2k} {\mathcal R}^4$ can lead to much more complex equations for the coefficient functions of the less fermionic interactions. This comes about because the less fermionic couplings mix with many couplings of different space-time structure each with its own set of modular forms.  Some of the interactions for a fixed space-time structure can therefore involve 
modular forms $z(\tau, \bar\tau)$ which split into
\be z(\tau, \bar\tau) = \sum_i z_i (\tau,\bar\tau) ,\ee
where each $z_i (\tau,\bar\tau)$ satisfies  a Poisson equation sourced by interactions in the effective action. In turn, each of these sources can involve more than one modular form leading to a highly  intricate, nested structure discussed more in section~\ref{implications}.

\section{Deriving the Recursion Relations}
\label{derivation}

We consider the space-time interactions in the low energy effective action given by 
\be \label{defchiral} {\rm det} e f^{(12+k,-12-k)} (\tau, \bar\tau)\hat{G}^{2k}\, \lambda^{16}, \ee
which we expect are related to  $D^{2k} {\mathcal{R}}^4$ as part of a supermultiplet of couplings obtained using supersymmetry along the lines described in section~\ref{sketch}. Here $f^{(12+k,-12-k)}$ is a modular form of weight 
$(12+k,-12 -k)$ defined in~\C{modformvar}. In principle, knowing the coupling~\C{defchiral}\ should be
enough to determine all the couplings of the supermultiplet using $SL(2,\mathbb{Z})$
invariance and supersymmetry. 

Since $\lambda^{16}$ forms a spacetime scalar, $\hat{G}^{2k}$ forms a scalar as well, and we
need to know the index contractions. Among the various possible interactions, we shall focus
on a particular contraction only, though we will discuss how the results generalize to the other cases. 
Couplings of the form $\hat{G}^{4g-4} {\mathcal{R}}^4$ have been discussed in the
literature~\cite{Berkovits:1994vy,Ooguri:1995cp,Berkovits:1998ex}. Our results differ from the earlier conjectures of~\cite{Berkovits:1998ex}. 

The structure $\hat{G}^{4g-4} {\mathcal{R}}^4$  gives  couplings 
${\mathcal{R}}^4$, $\hat{G}^4 {\mathcal{R}}^4$, $\hat{G}^8 {\mathcal{R}}^4 , \ldots$ but not ones of the 
form $\hat{G}^6 {\mathcal{R}}^4$, $\hat{G}^{10} {\mathcal{R}}^4 , \ldots$. However, it is easy to work out the specific space-time structure 
we want to study based on the details of the spacetime structure of the $\hat{G}^{4g-4} {\mathcal{R}}^4$ couplings, which we briefly sketch.
These couplings were first determined in six dimensions and then a Lorentz covariant expression for these interactions was obtained in eight dimensions of the form
\be \int d^8 x \Big( \Gamma^{\mu_1\nu_1\rho_1}_{a_1b_1} \ldots 
\Gamma^{\mu_4 \nu_4 \rho_4}_{a_4b_4} \delta^{[a_1}_{[b_1} \delta^{a_2}_{b_2} \delta^{a_3}_{b_3}
\delta^{a_4]}_{b_4]} H_{\mu_1\nu_1\rho_1} \ldots  H_{\mu_4\nu_4\rho_4} \Big)^{g-1} {\mathcal{R}}^4.\ee
Here $H_{\mu\nu\rho}$ is the RR $3$-form field strength obtained from ten dimensions. There is an additional  eight-dimensional field strength in the RR sector coming from $F_5$ which we drop. Among other terms, this gives
\be \int d^8 x \Big( (H_{\mu\nu\rho} H^{\mu\nu\rho})^2 (H_{\s\omega\xi} H^{\s\omega\xi})^2 \Big)^{g-1} 
{\mathcal{R}}^4 \sim \int d^8 x H^{4g-4}{\mathcal{R}}^4.\ee
Because the interactions $\hat{G}^6 {\mathcal{R}}^4$,  
$\hat{G}^{10} {\mathcal{R}}^4 , \ldots$ are also non--vanishing, we conclude that the combination
$H_{\mu\nu\rho} H^{\mu\nu\rho}$ must arise. $SL(2,\mathbb{Z})$ invariance and supercovariantizing 
then leads to the combination $\hat{G}_{\mu\nu\rho} \hat{G}^{\mu\nu\rho}$. We shall therefore focus on the structure $(\hat{G}_{\mu\nu\rho} \hat{G}^{\mu\nu\rho})^k$ in $\hat{G}^{2k}$ and drop
all other space-time structures. So
\be \hat{G}^{2k} \equiv (\hat{G}_{\mu\nu\rho} \hat{G}^{\mu\nu\rho})^k \ee
for our purposes.

Now consider the following interactions 
\be \label{threeterms}
\hat{G}^{2k} \lambda^{16}, \qquad \hat{G}^{2k+1} \lambda^{14}, \qquad \hat{G}^{2(k-1)} 
(\hat{G}^{\mu\nu\rho} \hat{G}_{\mu\nu\rho}^*) \lambda^{16},\ee
which are in the action $S^{(k+3)}$ at $O(\alpha'^{k+3})$ where we normalize the supergravity contribution to be $O(1)$ in $\alpha'$. Here
\be \hat{G}^{2k+1} \lambda^{14} \sim \hat{G}^{2k} \hat{G}^{\mu\nu\rho} (\g_{\mu\nu\rho} \g^0)_{ab} 
(\lambda^{14})^{ab}.\ee
These three interactions are special in the sense that they mix with each other under supersymmetry
but with no other interactions in the effective action at the same order in $\alpha'$.

\subsection{Sufficiently small $k$}

First we consider the case when $k<N$ is sufficiently small so that
\be \hat{G}^{2k} =  (-6i)^{2k} (\bar\psi^{*[\mu} \g^\nu \psi^{\rho]} 
\bar\psi^*_{[\mu} \g_\nu \psi_{\rho]} )^k + \ldots \equiv (-6i \psi \psi)^{2k} + \ldots.\ee

\subsubsection{Contributions from $L^{(k+3)}$ and $L^{(0)}$}
\label{sugracontributions}

Let us first consider the contributions only from $L^{(k+3)}$ and the supergravity action $L^{(0)}$. We shall consider
the source contributions from the terms in the effective action which are intermediate in orders of $\alpha'$ later.

Take the interactions from~\C{threeterms}, 
\bea \label{maincoup}
L^{(k+3)}_1 &&= {\rm det} e  f^{(12+k,-12-k)} (\tau,\bar\tau) (-6i \psi \psi)^{2k} \lambda^{16},\non \\
L^{(k+3)}_2 &&= {\rm det} e f^{(11+k,-11-k)} (\tau,\bar\tau) (-6i \psi \psi)^{2k} \lambda^{15} \g^\mu \psi_\mu^* ,
\non \\ L^{(k+3)}_3 &&= {\rm det} e \hat{f}^{(11+k,-11-k)} (\tau,\bar\tau) (-6i \psi \psi)^{2(k-1)}
(\hat{G}^{\mu\nu\rho} \hat{G}_{\mu\nu\rho}^*)\lambda^{16}, \eea
and consider their variations under the linearized supersymmetry transformation, $\delta^{(0)}$, given in Appendix~\ref{susysummary}, into 
$${\rm det} e 
(-6i \psi \psi)^{2k} \lambda^{16} (\bar\epsilon^* \g^\mu \psi_\mu^*).$$
We find that
\bea \delta^{(0)} L^{(k+3)}_1 &=& (\delta^{(0)} {\rm det} e \lambda^{16}) f^{(12+k,-12-k)} (\tau,\bar\tau) 
(-6i \psi \psi)^{2k} 
\non \\ &&
+ {\rm det} e f^{(12+k,-12-k)} (\tau,\bar\tau) \lambda^{16} (\delta^{(0)} \hat{G}^{2k}) +\ldots \non \\
&=& -8i {\rm det} e f^{(12+k,-12-k)} (\tau,\bar\tau) (-6i \psi \psi)^{2k} \lambda^{16}
(\bar\epsilon^* \g^\mu \psi_\mu^*) \non \\ &&
+ {\rm det} e f^{(12+k,-12-k)} (\tau,\bar\tau) \lambda^{16} (\delta^{(0)} \hat{G}^{2k}) 
+\ldots. \eea 
The last term receives two kinds of contributions: one is of the form $(\psi \psi)^{2(k-1)}
(\psi \psi) \cdot (\psi \delta^{(0)}\psi)$ given by 
the supervariation $\delta^{(0)} \psi_\mu$. This involves the piece of 
$D_\mu \epsilon$ of the form~\cite{Schwarz:1983qr}
\be D_\mu \epsilon = -\frac{i}{4} (\bar\psi_\nu \g_\mu \psi_\s + \bar\psi_\nu \g_\s \psi_\mu 
+\bar\psi_\mu \g_\nu \psi_\s) \g^{\nu\s} \epsilon +\ldots, \ee
and the $\psi^* \psi \epsilon$ term in $\hat{F}_5$. The other contribution is of the form
$(\psi \psi)^{2(k-1)} (\psi \psi) \cdot (\psi^* \delta^{(0)} \lambda)$ using 
$$\delta^{(0)} \lambda
\sim \epsilon \hat{G} \sim \epsilon \psi \psi,$$ 
and gives a contribution of the required type after using a Fierz identity.
The precise numerical values are not relevant for our purposes, and we finally find that
\be \delta^{(0)} L^{(k+3)}_1 \sim
{\rm det} e f^{(12+k,-12-k)} (\tau,\bar\tau) (-6i \psi \psi)^{2k} \lambda^{16}
(\bar\epsilon^* \g^\mu \psi_\mu^*) + \ldots.\ee
We also find that
\bea \delta^{(0)} L^{(k+3)}_2 &=& {\rm det} e \delta^{(0)} (f^{(11+k,-11-k)} (\tau,\bar\tau) \lambda^{15} \g^\mu 
\psi_\mu^*) (-6i \psi \psi)^{2k}\non \\ &&+ {\rm det} e f^{(11+k,-11-k)} (\tau,\bar\tau) \lambda^{15} \g^\mu \psi_\mu^* 
\delta^{(0)}((-6i \psi \psi)^{2k}) + \ldots\non \\ &=& 2i {\rm det} e (-6i \psi \psi)^{2k} \lambda^{16}
(\bar\epsilon^* \g^\mu \psi_\mu^*) D_{11} f^{(11+k,-11-k)} (\tau,\bar\tau)\non \\ &&+ 
{\rm det} e f^{(11+k,-11-k)} (\tau,\bar\tau) \lambda^{15} \g^\mu \psi_\mu^* \delta^{(0)}((-6i \psi \psi)^{2k})+ \ldots. \eea
The second term receives a contribution from $\delta^{(0)} \psi \sim \psi \lambda \epsilon$, leading to
\be \delta^{(0)} L^{(k+3)}_2 =2i {\rm det} e (-6i \psi \psi)^{2k} \lambda^{16}
(\bar\epsilon^* \g^\mu \psi_\mu^*) D_{11+k} f^{(11+k,-11-k)} (\tau,\bar\tau) + \ldots.\ee
We have kept track of the factors to show the emergence of the modular covariant derivative with the correct 
modular weight. We finally also see that
\be \delta^{(0)} L^{(k+3)}_3 =0+ \ldots.\ee
Thus
\bea \label{impvarone}
&& \delta^{(0)} (L^{(k+3)}_1 + L^{(k+3)}_2 +L^{(k+3)}_3) \sim  i \Big( D_{k +11} f^{(11+k,-11-k)} (\tau,\bar\tau) 
\non \\ &&+  f^{(12+k,-12-k)} (\tau,\bar\tau) \Big)
{\rm det} e (-6i \psi \psi)^{2k} \lambda^{16}
(\bar\epsilon^* \g^\mu \psi_\mu^*) + \ldots. \eea

Next we consider the variations under linearized supersymmetry into 
$$({\rm det} e) (-6i \psi \psi)^{2k} \lambda^{16} (\bar\epsilon \lambda^*).$$
We find that
\bea \delta^{(0)} L^{(k+3)}_1 &=& {\rm det} e \delta^{(0)} ( f^{(12+k,-12-k)} (\tau,\bar\tau)\lambda^{16})
(-6i\psi\psi)^{2k} 
\non \\ &&+ {\rm det} e f^{(12+k,-12-k)} (\tau,\bar\tau) \lambda^{16} \delta^{(0)}((-6i \psi \psi)^{2k})  
+ \ldots
\non \\ &=& -2i {\rm det} e (-6i \psi \psi)^{2k} \lambda^{16} (\bar\epsilon \lambda^*) 
\bar{D}_{-12} f^{(12+k,-12-k)} (\tau,\bar\tau) 
\non \\ &&
+ {\rm det} e f^{(12+k,-12-k)} (\tau,\bar\tau) \lambda^{16} \delta^{(0)}((-6i \psi \psi)^{2k})+ \ldots  
\non \\
&=& -2i {\rm det} e (-6i \psi \psi)^{2k} \lambda^{16} (\bar\epsilon \lambda^*) 
\bar{D}_{-(12 +k)} f^{(12+k,-12-k)} (\tau,\bar\tau)+ \ldots, \eea
where for the last term we have used $\delta^{(0)}\psi \sim \psi \lambda^* \epsilon^*$. Again we get the
modular covariant derivative with the correct modular weight. 
Also, we see that
\bea \delta^{(0)} L^{(k+3)}_2 &=& {\rm det} e f^{(11+k,-11-k)} (\tau,\bar\tau) (-6i \psi \psi)^{2k} 
\lambda^{15} \delta^{(0)} (\g^\mu \psi_\mu^*) + \ldots
\non \\ &\sim& i {\rm det} e f^{(11+k,-11-k)} (\tau,\bar\tau) 
(-6i \psi \psi)^{2k}\lambda^{16} (\bar\epsilon \lambda^*)+ \ldots,
\eea
where we have used~\cite{Green:1998by}  
\be \lambda^{15} \delta^{(0)} (\g^\mu \psi_\mu^*) = 15 i \lambda^{16} (\bar\epsilon \lambda^*).\ee
Finally, 
\bea \label{manyterms} \delta^{(0)} L^{(k+3)}_3 \sim i {\rm det} e \hat{f}^{(11+k,-11-k)} (\tau,\bar\tau) 
(-6i \psi \psi)^{2k}\lambda^{16} (\bar\epsilon \lambda^*)+ \ldots.\eea
In the calculation of \C{manyterms}, the relevant supervariations that give us the required spacetime structure
involve 
$$\delta^{(0)} e \sim \epsilon^* \psi, \quad \delta^{(0)} \lambda \sim \epsilon^* \hat{P} \sim \epsilon^*
\psi \lambda, \quad \delta^{(0)} \psi \sim \hat{G} \epsilon^*, $$ 
$$ \delta^{(0)} \hat{G} \sim \psi \delta^{(0)} \psi \sim \psi \hat{G} 
\epsilon^* \sim \psi^3 \epsilon^*,$$ and 
$$\delta^{(0)} \hat{G}^*
\sim (\delta^{(0)} \psi) \lambda^* + \psi (\delta^{(0)} \lambda^*) \sim \hat{G} \epsilon^* \lambda^* +
\psi \epsilon^* \hat{G}^* \sim \psi \psi \epsilon^* \lambda^*. $$ 
To show that they give the spacetime structure
in \C{manyterms}, one has to use the Fierz identity extensively. To summarize: 
\bea \label{impvartwo}
&& \delta^{(0)} (L^{(k+3)}_1 + L^{(k+3)}_2 +L^{(k+3)}_3) \sim i \Big( \bar{D}_{-(k +12)} f^{(12+k,-12-k)} (\tau,\bar\tau) 
\non \\ 
&& +  f^{(11+k,-11-k)} (\tau,\bar\tau) + \hat{f}^{(11+k,-11-k)} (\tau,\bar\tau)\Big)
{\rm det} e (-6i \psi \psi)^{2k} \lambda^{16}
(\bar\epsilon \lambda^*) + \ldots. \eea

Let us now consider the contributions from supergravity. First consider contributions from $\delta^{(k+3)}
L^{(0)}$ which give us $(-6i\psi \psi)^{2k} \lambda^{16} (\bar\epsilon \lambda^*)$. Let us start with the $\lambda^{*2} \lambda^2$ term in the supergravity action 
given by (ignoring an overall coefficient of $1/256$)~\cite{Green:1998by} 
\be L^{(0)}_1 =
{\rm det} e (\bar\lambda^* \g^{\mu\nu\rho} \lambda )(\bar\lambda\g_{\mu\nu\rho} \lambda^*) .
\footnote{Note that $(\bar\lambda^* \g^{\mu\nu\rho} \lambda )(\bar\lambda\g_{\mu\nu\rho} \lambda^*)
= 6 (\bar\lambda^* \g^\mu \lambda )(\bar\lambda \g_\mu \lambda^*)$ by Fierzing, so this expression
is the unique one.}\ee
Note that this term is not obtained from the $\hat{F}_5^2$ term in the action because of the spinor identity
\C{vanspin}, and has to be constructed separately. Consider the set of supervariations given by
\bea \label{partvar}
\delta^{(k+3)} \lambda^*_a &=& i (-6i\psi \psi)^{2k} \Big[g_1 (\tau, \bar\tau) (\lambda^{14})_{cd} 
(\g^{\mu\nu\rho} \g^0)_{dc} (\g_{\mu\nu\rho} \epsilon^*)_a  + g_2 (\tau, \bar\tau)
(\lambda^{14})_{ab} (\g^0 \epsilon^*)_b \non \\ &&+ (\lambda^{14})_{bd}
\Big( g_3 (\tau, \bar\tau) (\g^0)_{ab} \epsilon^*_d + g_4 (\tau, \bar\tau)
(\g^\mu \g^0)_{ab} (\g_\mu \epsilon^*)_d  \\ &&+ g_5 (\tau, \bar\tau)
(\g^{\mu\nu} \g^0)_{ab} (\g_{\mu\nu} \epsilon^*)_d + g_6 (\tau, \bar\tau)
(\g^{\mu\nu\rho} \g^0)_{ab} (\g_{\mu\nu\rho} \epsilon^*)_d  \non \\ &&+g_7 (\tau, \bar\tau)
(\g^{\mu_1 \cdots \mu_4} \g^0)_{ab} (\g_{\mu_1 \cdots \mu_4} \epsilon^*)_d 
+g_8 (\tau, \bar\tau) (\g^{\mu_1 \cdots \mu_5} \g^0)_{ab} (\g_{\mu_1 \cdots \mu_5} \epsilon^*)_d \Big)\Big] , \non\eea
which are all the possible supervariations that survive for $k=0$. We set $g_6 =0$ by redefining
\be g_1 \rightarrow g_1 + \frac{g_6}{2} ,\ee
and using the identity
\be \label{iden10}
(\g_{\mu\nu\rho} \g^0)_{dc} \g^{\mu\nu\rho}_{ab} 
- (\g_{\mu\nu\rho} \g^0)_{ac} \g^{\mu\nu\rho}_{db} 
+ (\g_{\mu\nu\rho} \g^0)_{ad} \g^{\mu\nu\rho}_{cb}=0.\ee
The identity~\C{iden10} can be proved by noticing that the expression is antisymmetric in $(c,d)$ and
thinking of $(a,b)$ as irrelevant indices. Thus constraints of chirality and antisymmetry force it to be 
proportional to $(\g_{\mu\nu\rho} \g^0)_{dc}$. Note that $(\g_\mu \g^0)_{dc}$ and 
$(\g_{\mu_1 \cdots \mu_5} \g^0)_{dc}$ are symmetric in $(c,d)$. This immediately leads to~\C{iden10}\ after multiplying by
$(\g^0 \g_{\s_1 \s_2 \s_3})_{cd}$.~\footnote{This proof is along the lines of Appendix 4.A in~\cite{GSW1}.} 

Now~\C{partvar}\ gives us the relation
\bea \label{impvarthree}
\frac{1}{18 \cdot 32} \delta^{(k+3)} L^{(0)}_1 = i(-6i \psi \psi)^{2k} \hat{g} (\tau, \bar\tau) 
{\rm det} e \lambda^{16} (\bar\epsilon \lambda^*),\eea
where
\bea \hat{g} (\tau, \bar\tau) &=& 
-480 g_1(\tau, \bar\tau) - 2 g_2 (\tau, \bar\tau)+ 5 g_3 (\tau, \bar\tau)- 20 g_4 (\tau, \bar\tau)
\non \\ &&- 30 g_5 (\tau, \bar\tau) -1680 g_7(\tau, \bar\tau).\eea
Note that $\hat{g} (\tau, \bar\tau)$ is independent of $g_8 (\tau, \bar\tau)$. 

We now impose the constraint of closure of the supersymmetry algebra to vastly reduce the number
of coefficients in \C{partvar}. Since we do not have an off-shell superspace formalism, the supersymmetry algebra
closes only with the use of the equations of motion for the fermionic fields, modulo various local symmetry transformations  of the theory.

We begin by considering
\be \delta = \delta^{(0)} + \alpha'^{k+3} \delta^{(k+3)} ,\ee
and restrict only to the part of $[\delta_1 , \delta_2] \lambda^*$ that depends on $\epsilon_1$ and $\epsilon_2^*$. 
The commutator of two supersymmetry transformations gives, 
\be [\delta_1,\delta_2] \lambda^* = \left([\delta^{(0)}_1,\delta^{(0)}_2] +\alpha'^{k+3}[\delta^{(0)}_1,\delta^{(k+3)}_2]
+ \alpha'^{k+3}[\delta^{(k+3)}_1,\delta^{(0)}_2] \right) \lambda^* + \ldots .\ee
The supergravity contribution is given by~\cite{Schwarz:1983qr}
\be \label{sugraclos}
[\delta^{(0)}_1,\delta^{(0)}_2] \lambda^*_a = \xi^\mu D_\mu \lambda^*_a 
+ i \Big( -\frac{3}{8} (\bar\epsilon_2 \g_\nu \epsilon_1)
\g^\nu_{ab} + \frac{1}{96} (\bar\epsilon_2 \g_{\nu_1 \nu_2 \nu_3} \epsilon_1) \g^{\nu_1 \nu_2 \nu_3}_{ab} 
\Big) (\g^\mu D_\mu \lambda^*)_b + \ldots ,\ee
where
\be \xi^\mu = i \bar\epsilon_2 \g^\mu \epsilon_1 .\ee
We see that closure follows after using the free equation of motion. 
The specific space-time structure in \C{sugraclos} is crucial in determining the higher 
derivative corrections. 

Let us next consider $[\delta^{(0)}_1,\delta^{(k+3)}_2] \lambda^*$. Keeping only the
relevant terms, we find
\bea \label{leadord}
[\delta^{(0)}_1,\delta^{(k+3)}_2] \lambda^*_a &=& 192 (-6i \psi \psi)^{2k} (\lambda^{15})_c
\Big( -\frac{3}{8} (\bar\epsilon_2 \g_\nu \epsilon_1)
\g^\nu_{ab} + \frac{1}{96} (\bar\epsilon_2 \g_{\nu_1 \nu_2 \nu_3} \epsilon_1) \g^{\nu_1 \nu_2 \nu_3}_{ab} 
\Big) \g^0_{bc} (D_{11+k} g_1)  
\non \\  && - 6 ik (-6i \psi \psi)^{2(k-1)} g_1 (\lambda^{14})_{cd} 
(\g^{\mu\nu\rho} \g^0)_{dc} (\g_{\mu\nu\rho} \epsilon^*_2)_a  \hat{G}_{\s_1\s_2 \s_3} 
((\delta^{(0)}_1\bar\psi^{\s_1}) \g^{\s_2 \s_3} \lambda )
\non  \\ && +
\frac{1}{4} (-6 i \psi \psi)^{2k} g_1  (\bar\epsilon^*_1 \lambda)
(\lambda^{14})_{cd} (\g^{\mu\nu\rho} \g^0)_{dc} (\g_{\mu\nu\rho} \epsilon^*_2)_a , \eea
where we have used the Fierz identity. The second term in \C{leadord} gives a contribution different from
the others which we discuss later. 

In calculating the first term, we see that the 
contributions from the other $g_i$ functions do not give the space-time
structure appearing in~\C{sugraclos}, and so they vanish. The contributions from $g_2, g_3, g_5, g_7$, and  $g_8$ 
involve a term with $5$ gamma matrices of the form
\be (\lambda^{15})_b (\bar\epsilon_2 \g_{\mu_1 \cdots \mu_5} \epsilon_1) (\g^{\mu_1 \cdots \mu_5}
\g^0)_{ab}, \ee
while \C{sugraclos} has no more than 3 gamma matrices, and so they vanish. The contribution from $g_4$ 
gives the spacetime structure
\be  -\frac{9}{8} (\bar\epsilon_2 \g_\nu \epsilon_1)
\g^\nu_{ab} + \frac{1}{24} (\bar\epsilon_2 \g_{\nu_1 \nu_2 \nu_3} \epsilon_1) \g^{\nu_1 \nu_2 \nu_3}_{ab} \ee
and so $g_4 =0$. Simply the space-time structure of~\C{sugraclos}\ therefore allows only one
non-vanishing coefficient in~\C{partvar}. The last term in~\C{leadord}\ is a supersymmetry transformation
of the type~\C{partvar}\ with
\be \epsilon = -\frac{i}{4} (\bar\lambda \epsilon^*_1) \epsilon_2 .\ee

The second term in~\C{leadord}\ gets a non-trivial contribution from $\delta^{(0)}_1 \psi^* \sim \epsilon_1 \hat{G}^*$ and is proportional to
\be ik (-6i \psi \psi)^{2(k-1)} (\hat{G}^{\mu\nu\rho} \hat{G}_{\mu\nu\rho}^*) g_1  
(\lambda^{15})_c
\Big( -\frac{3}{8} (\bar\epsilon_2 \g_\nu \epsilon_1)
\g^\nu_{ab} + \frac{1}{96} (\bar\epsilon_2 \g_{\nu_1 \nu_2 \nu_3} \epsilon_1) \g^{\nu_1 \nu_2 \nu_3}_{ab} 
\Big) \g^0_{bc}.\ee
We see that
\be [\delta^{(k+3)}_1,\delta^{(0)}_2] \lambda^*_a = 0 \ee
for this type of contribution. 
Thus closure of the supersymmetry algebra for $\lambda^*$ gives the dilatino equation of motion
\bea \label{impvarfour}
 & \g^0 \g^\mu D_\mu \lambda^* &+  \alpha'^{k+3} (D_{k+11} g_1 )(-6i \psi \psi)^{2k} \lambda^{15}
\cr &&+   \alpha'^{k+3} g_1 (-6i \psi \psi)^{2(k-1)} (\hat{G}^{\mu\nu\rho} \hat{G}_{\mu\nu\rho}^*) \lambda^{15}
+\ldots =0.\eea

We began by considering only those supervariations which survive for $k=0$. 
We can also have other supervariations
which contribute only for $k \geq 2$. For example, we can have something like 
\be \delta^{(k+3)} \lambda^*_a \sim (\psi \psi)^{2(k-1)} (\bar\psi^{*[\rho_1} \g^{\tau_1} 
\psi^{\tau_2]})(\bar\psi^*_{[\rho_2} \g_{\tau_1} \psi_{\tau_2]}) (\lambda^{14})_{cd} 
(\g_{\rho_1}^{~\s_1 \s_2} \g^0)_{dc} (\g^{\rho_2}_{~\s_1 \s_2} \epsilon^*)_a ,\ee
among many other possibilities. However, arguing along the same lines as above, we find that there are actually (and remarkably) no other terms.

Next we consider the contribution from the $\lambda^* \psi \psi \psi$ part of the action, which comes
from expanding the $\hat{G} \cdot \hat{G}^*$ coupling. There are two contributions of the form $\lambda^* \psi \psi (\delta^{(k+3)
} \psi)$ and $(\delta^{(k+3)} \lambda^*) \psi \psi \psi$.  

Let us first study contributions from $\lambda^* \psi \psi (\delta^{(k+3)} \psi)$. Dropping an irrelevant 
numerical factor, consider the term in the Lagrangian
\be L^{(0)}_2 = -6i {\rm det} e (\bar\psi^*_{[\mu} \g_\nu \psi_{\rho]}) (\bar\lambda \g^{[\mu\nu} \psi^{\rho]}).\ee
Supercovariance of the theory allows the possible super-variation
\be \delta^{(k+3)} \psi_\mu^a = i (-6i \psi \psi)^{2(k-1)} \lambda^{16} \Big( p_1 (\tau, \bar\tau) 
(\g^{\s_1 \s_2 \s_3} \g_\mu \epsilon^*)^a + p_2 (\tau, \bar\tau) (\g_\mu \g^{\s_1 \s_2 \s_3}  \epsilon^*)^a
\Big) \hat{G}_{\s_1\s_2 \s_3} , \non \ee
which, after using  
\be 
\delta^{(k+3)} L^{(0)}_2 = -6i {\rm det} e \Big( (\bar\lambda \g^{[\mu\nu} \delta^{(k+3)} \psi^{\rho]})
(\bar\psi^*_{[\mu} \g_\nu \psi_{\rho]}) + 2 (\bar\psi^*_{[\mu} \g_\nu \delta^{(k+3)} \psi_{\rho]})
(\bar\lambda \g^{[\mu\nu} \psi^{\rho]}) \Big),\ee
gives 
\be \label{impvarfive}
\delta^{(k+3)} L^{(0)}_2 = i(-6i \psi \psi)^{2k} \hat{p} (\tau, \bar\tau) 
{\rm det} e \lambda^{16} (\bar\epsilon \lambda^*),\ee
where
\be \hat{p} (\tau, \bar\tau) \sim p_1 (\tau, \bar\tau) + p_2 (\tau, \bar\tau),\ee
after extensive use of the Fierz identity and the relation
\bea \hat{G}_{\mu\nu\rho} \hat{G}_{\rho_1 \rho_2 \rho_3} \bar\lambda \g^{\mu\nu} \g^{\rho_1 \rho_2 \rho_3} \g^\rho
\epsilon^* = -6 \hat{G}^2 \bar\lambda \epsilon^* + \ldots
.\eea
Using the symmetry under interchange of $p_1$ and $p_2$, we can set $p_2 =0$. In fact, we shall only use the closure of $\g^\mu 
\psi_\mu$, where the two contributions are proportional.

We now use the closure of the supersymmetry algebra on $ \g^\mu \psi_\mu$ to constrain 
$p_1$ (which is easier to calculate than the closure of $\psi_\mu$). 
First consider the closure at the level of supergravity where we keep only the terms
proportional to $\epsilon_1 \epsilon^*_2 D \psi$ in our analysis. We make use of the 
definitions~\cite{Schwarz:1983qr}\footnote{The term proportional to $Q_\mu \epsilon$
in $\delta^{(0)} \psi_\mu$ is not needed because $\delta^{(0)} V^\alpha_+ = V^\alpha_- \bar\epsilon^* \lambda$, and does 
not vary into a gravitino. We shall consider the additional supervariation to $G_{\mu\nu\rho}$ coming
from the compensating $U(1)$ gauge transformation later. It is not relevant for the present analysis.}
\bea \label{impdef}
F^\alpha_{\mu\nu\rho} =3 \p_{[\mu} A^\alpha_{\nu\rho]}, 
\qquad G_{\mu\nu\rho} = -\epsilon_{\alpha \beta} V^\alpha_+ F^\beta_{\mu\nu\rho},~~~~~~~~ \non \\
D_\mu \epsilon = \frac{1}{4} \Big( e_\mu^m \p_{[\nu} e_{\rho]m} + e_\rho^m \p_{[\nu} e_{\mu]m}
+e_\nu^m \p_{[\mu} e_{\rho]m} \Big) \g^{\nu\rho} \epsilon + \ldots ,\eea
and the supersymmetry variation
\bea \label{addtodef}
\delta^{(0)} A_{\mu \nu}^\alpha = V^\alpha_+ \bar\epsilon^* \g_{\mu\nu} \lambda^* + V^\alpha_- \bar\epsilon
\g_{\mu\nu} \lambda + 4i V^\alpha_+ \bar\epsilon \g_{[\mu} \psi^*_{\nu]} + 4i V^\alpha_- 
\bar\epsilon^* \g_{[\mu} \psi_{\nu]}\eea
where $\alpha, \beta$ are the global $SU(1,1)$ indices. We have also used
\be -\epsilon_{\alpha \beta} V^\alpha_+ V^\beta_- =1.\ee
This gives us 
\bea [\delta^{(0)}_1, \delta^{(0)}_2] (\g^\mu \psi_\mu)^a &=& 
-\frac{i}{4} (\g^{\mu\nu\rho}  \epsilon_2^*)^a
(\bar\epsilon_1^* \g_\mu D_\nu \psi_\rho ) \non \\ && - \frac{i}{4} 
\bar\epsilon_2 \Big( \g_\mu D_{[\nu} \psi_{\rho]} + \g_\rho D_{[\nu} \psi_{\mu]}
+\g_\nu D_{[\mu} \psi_{\rho]} \Big) (\g^\mu\g^{\nu\rho} \epsilon_1)^a \eea
since the term involving $F_5$ vanishes using $\g^\mu \g_{\mu_1 \cdots \mu_5} \g_\mu =0$. 
Using the Fierz identity again, we obtain 
\bea \label{fierzclose}
[\delta^{(0)}_1, \delta^{(0)}_2] (\g^\mu \psi_\mu)^a = i (\bar\epsilon_2
\g^\s \epsilon_1) \Big( \frac{1}{4} \g_\s \g^{\mu\nu} D_\mu \psi_\nu 
+ \frac{1}{2} (\g^\mu D_\mu \psi_\s - \g^\mu D_\s \psi_\mu) \Big)^a \non \\
+\frac{i}{96} (\bar\epsilon_2
\g^{\s_1 \s_2 \s_3} \epsilon_1) \Big( -\g_{\s_1 \s_2\s_3} \g^{\mu\nu} D_\mu \psi_\nu
+ 3 \g_{\s_1 \s_2} (\g^\mu D_{\s_3} \psi_\mu - \g^\mu D_\mu \psi_{\s_3}) \Big)^a . \eea
Note that there is no term involving $(\bar\epsilon_2\g^{\s_1 \ldots \s_5} \epsilon_1)$.

For our purposes, it is enough to consider
\be \label{fierzclose2}
[\delta^{(0)}_1, \delta^{(0)}_2] (\g^\mu \psi_\mu)_a = \frac{i}{4}  \Big(
(\bar\epsilon_2 \g_\s \epsilon_1) \g^\s_{ab} - \frac{1}{24}
(\bar\epsilon_2 \g_{\s_1 \s_2 \s_3} \epsilon_1) \g^{\s_1 \s_2\s_3}_{ab}
\Big) (\g^{\mu\nu} D_\mu \psi_\nu)^b. \ee
The remaining terms in \C{fierzclose} are obtained from the closure of $[\delta^{(0)}_1, \delta^{(0)}_2] \psi_\mu$
by acting with $\g^\mu$. They involve contributions from the equation of motion in 
$[\delta^{(0)}_1, \delta^{(0)}_2] \psi_\mu$ as well as from the transformations
\be [\delta^{(0)}_1, \delta^{(0)}_2] \psi_\mu = \xi^\nu D_\nu \psi_\mu + D_\mu \tilde\epsilon 
+ \ldots,\ee 
where
\be \tilde \epsilon = - \psi_\nu \xi^\nu + \ldots, \ee
leading to (see section 1.9 of~\cite{VanNieuwenhuizen:1981ae} for a relevant discussion)
\be [\delta^{(0)}_1, \delta^{(0)}_2] \psi_\mu = \xi^\nu (D_\nu \psi_\mu - D_\mu \psi_\nu) + \ldots .\ee
These are the only local symmetry transformations appearing in the closure that involve $D \psi$. They correspond to
general coordinate transformations and supersymmetry transformations, respectively. 
Thus the $\bar\epsilon_2 \g^{\s_1\s_2 \s_3} \epsilon_1$ term
in~\C{fierzclose}\ receives
contributions only from the equation of motion. However, the $\bar\epsilon_2 \g^\s \epsilon_1$ term 
in~\C{fierzclose}\ receives contributions from both the equations of motion as well as from the
local symmetry transformations.     

Proceeding as before, we can calculate the closure involving the higher derivative corrections. This gives,
\bea \label{psiclos}
&&([\delta^{(0)}_1, \delta^{(k+3)}_2 ] + [\delta^{(k+3)}_1, \delta^{(0)}_2]) (\g^\mu \psi_\mu)^a
 \sim \cr && (p_1 + g_1) \hat{G}^{2k} (\lambda^{15})^c \Big( (\bar\epsilon_2 \g_\s \epsilon_1) \g^\s_{ab}
  + (\bar\epsilon_2 \g_{\s_1 \s_2 \s_3} \epsilon_1) \g^{\s_1 \s_2\s_3}_{ab} \Big) \g^0_{bc} . \eea
In \C{psiclos}, the relevant supergravity transformations involving $p_1$
are given by $\delta^{(0)} \lambda \sim \epsilon \hat{G}$ and $\delta^{(0)} \psi \sim \psi^* \psi \epsilon$,
while those involving $g_1$ are given by $\delta^{(0)}\psi \sim \lambda^* \lambda \epsilon$. Some of
them give the required terms directly, while the rest give the required terms on Fierzing
between $\psi^*$ and $\lambda$.   

Note that unlike the previous case, we do not need to work out the specific coefficients in the closure in
\C{psiclos}. Earlier, even though we are not interested in the exact coefficients, we needed the exact coefficients
in the closure to eliminate all but one coefficient in $\delta^{(k+3)} \lambda^*$. However here we have only one
coefficient to begin with so it is good enough to show that the expected terms in the supergravity closure
(and no others) arise in the closure involving higher derivatives.\footnote{This 
will be the strategy followed later on as well. It could be 
that the specific space-time structure does not match, in which case, the corresponding coefficient vanishes.}   

Therefore closure of the supersymmetry algebra for $\psi$ gives the gravitino equation of motion
\be \label{impvarsix}
\g^0 \g^{\mu\nu\rho} D_\nu \psi_\rho + \alpha'^{(k+3)} (p_1 + g_1 )(-6i \psi \psi)^{2k} \g^\mu \lambda^{15}
+\ldots =0.\ee

Next consider the contribution from $(\delta^{(k+3)} \lambda^*) \psi \psi \psi$. The possible supervariations
are given by
\be \label{notposs}
\delta^{(k+3)} \lambda^*_a =  (\lambda^{15})_b \hat{G}^{2(k-1)} (\bar\lambda \epsilon^*)
(\bar\psi^*_\mu \Gamma^M \lambda) (\g^\mu \Gamma_M \g^0)_{ab}, \ee
where $\Gamma^M = \{1, \g^{\mu\nu} , \g^{\mu_1 \cdots \mu_4}\}$.
Note that the only supercovariant supervariation has $\Gamma^M =1$,
because $\hat{P}_\mu \sim \bar\psi^*_\mu \lambda$. 
However, every term in~\C{notposs}\ is actually proportional to 
$$\hat{G}^{2(k-1)}(\bar\lambda \epsilon^*) \lambda^{16} (\g^\mu \psi_\mu)_a,$$ 
and is inconsistent with the closure of the 
superalgebra \C{sugraclos}.
Hence there are no such
contributions.

There are no contributions from the $(\psi \psi^*)^2$ and $\lambda (\psi\psi \psi)^*$ terms in the action, The only remaining possibility is a contribution from the $\lambda \lambda^* \psi \psi^*$ term in the action
given by $$(\delta^{(k+3)} \psi^*) \lambda^* \lambda \psi.$$
This comes from the $\hat{P} \cdot \hat{P}^*$, $\hat{G} \cdot \hat{G}^*$, and $F_5^2$ terms in the action. Thus
\be L^{(0)}_3  \sim {\rm det} e [(\bar\psi^*_\mu \lambda) (\bar\lambda \psi^{\mu *})+ (\bar\lambda
\g^{[\mu\nu} \psi^{\rho]}) (\bar\psi_\mu \g_{\nu\rho} \lambda) + (\bar\lambda
\g^{\mu_1 \cdots \mu_5} \lambda) (\bar\psi_{\mu_1} \g_{\mu_2 \mu_3 \mu_4} \psi_{\mu_5})]. \ee
It is not difficult to construct $\delta^{(k+3)}\psi^*$ such that 
$$\delta^{(k+3)} L^{(0)}_3 \sim {\rm det}e (-6i\psi \psi)^{2k}
\lambda^{16} (\bar\epsilon \lambda^*).$$ 
For example, we can take
\bea \delta^{(k+3)} \psi^{a*}_\mu &\sim & (\lambda^{15})^a \hat{G}^{2(k-1)} \hat{G}_{\mu\nu\rho} (\epsilon^* \g^\nu\psi^\rho)
+ (\lambda^{15})^b \hat{G}^{2(k-1)}\hat{G}_\mu^{~\rho_1\rho_2} (\bar\psi^*_{\rho_1} \g_{\rho_2} \g^0)_b \epsilon^{a*}
\non \\ &&
+(\lambda^{15})^b \hat{G}^{2(k-1)}\hat{G}_\mu^{~\rho_1\rho_2} (\bar\psi^*_{\rho_1} \g_{\rho_2} \g^0)^a \epsilon_{b*}, \eea
where every term is multiplied by a modular form. However, no supervariation of $\psi^*$ is consistent with the closure
of the supersymmetry algebra given by \C{fierzclose2}. This is because the contribution of the type $(\bar\epsilon_2
\g^{\s_1 \cdots \s_5} \epsilon_1)$ to $([\delta^{(0)}_1, \delta^{(k+3)}_2]+[\delta^{(k+3)}_1,\delta^{(0)}_2])\psi^*$ is 
nonvanishing which contradicts~\C{fierzclose2}. From now on, we will list only non-trivial supervariations.

Next consider contributions from $\delta^{(k+3)} L^{(0)}$ which give 
$$(-6i\psi \psi)^{2k} \lambda^{16} (\bar\epsilon^* \g^\mu \psi_\mu^*).$$ 
There are no supervariations $\delta^{(k+3)} \psi^*$ 
or $\delta^{(k+3)} \lambda^*$ which give
the required $\delta^{(k+3)} L^{(0)}$ terms and are consistent with the closure of the superalgebra. 

If this were the complete analysis, from~\C{impvarone},~\C{impvartwo},~\C{impvarthree},~\C{impvarfour},~\C{impvarfive} and~\C{impvarsix}, it would follow that 
\be g_1 \sim p_1 \sim f^{(11+k, -11-k)} \sim \hat{f}^{(11+k, -11-k)} ,\ee
and that
\be D_{k+11} f^{(11+k, -11-k)} \sim f^{(12+k, -12-k)} , \qquad \bar{D}_{-(12+k)} f^{(12+k, -12-k)}
\sim f^{(11+k, -11-k)} ,\ee
leading to
\bea \label{Lapeqn}
\bar{D}_{-(12+k)} D_{k+11} f^{(11+k, -11-k)} &\sim & f^{(11+k, -11-k)}, \cr 
D_{k+11} \bar{D}_{-(12+k)} f^{(12+k, -12-k)} &\sim & f^{(12+k, -12-k)}.\eea
The couplings would satisfy Laplace equations on the fundamental domain of $SL(2,\mathbb{Z})$.

What is missing from the discussion are the source terms. So we next consider
the contributions coming from terms in the effective action which are intermediate orders 
in $\alpha'$. These sources correct the Laplace equations to Poisson equations. 
The basic idea is to further use the constraints coming from supersymmetry
\be \Big( \delta^{(0)} + \sum_{k=0}^\infty \alpha'^{k+3}\delta^{(k+3)} \Big) \Big(S^{(0)} + \sum_{k=0}^\infty \alpha'^{k+3}
S^{(k+3)} \Big) =0.\ee
Apart from the invariance of the supergravity action, the existence of the supersymmetry implies 
\be \label{sourceeqn}
\delta^{(0)} S^{(k+3)} + \delta^{(k+3)} S^{(0)} + \sum_{m> 0, n>0, m+n = k+3} \delta^{(m)} S^{(n)} = 0.\ee 
Only for $k=0$ and $2$ does the last term in~\C{sourceeqn}, which contains the source terms, not contribute. For those special cases,~\C{Lapeqn}\ gives the complete answer. 
The remaining equations all receive contributions from the source terms to which we will turn later.

\subsubsection{The specific cases of $k=0$ and $k=2$}

We now consider the equations in \C{Lapeqn} for $k=0$ and $2$. The constants of proportionality can be completely fixed using the expression for the four graviton amplitude at genus 
zero~\cite{Green:1981xx,DHoker:1988ta},
\bea \label{treeamp}
\mathcal{A}^{{\rm tree}} (s,t,u)&=& -\tau_2^2 \frac{\Gamma(-\alpha' s/4) \Gamma (-\alpha' t/4) \Gamma 
(-\alpha' u/4)}{\Gamma(1 + \alpha' s/4) \Gamma (1 + \alpha' t/4) \Gamma 
(1 + \alpha' u/4)} \mathcal{R}^4 \non \\ &&
= \tau_2^2 \Big( \frac{64}{\alpha'^3 stu} + 2 \zeta (3) + \frac{\zeta (5)}{16} 
\alpha'^2 (s^2 + t^2 + u^2) + \frac{\zeta (3)^2}{96} \alpha'^3 
(s^3 + t^3 + u^3) \non \\ &&+ \frac{\zeta (7)}{512} \alpha'^4 (s^2 + t^2 + u^2)^2 + 
\frac{\zeta (3) \zeta (5)}{1280} \alpha'^5 (s^5 + t^5 + u^5) +\ldots \Big)\mathcal{R}^4 ,\eea
where $s,t,$ and $u$ are the Mandelstam variables and $s+t+ u=0$.  

Let us consider the $k=0$ case first. From \C{treeamp}, we see that the 
$\mathcal{R}^4$ interaction has a tree level contribution proportional to $\zeta (3) \tau_2^{3/2}$ in 
Einstein frame where the metric is duality invariant. 
Also it can be shown that the genus one amplitude has a power law dependence in 
$\tau_2$ (this is also true for the genus two amplitude for the $k=2$ case which we discuss next). 
Because this has a unique space-time structure along with $\lambda^{15} \g^\mu \psi_\mu^*$, 
it follows that the tree-level contribution to 
$$f^{(11,-11)} \sim D_{10} \ldots D_0 \zeta (3) \tau_2^{3/2} \sim
\zeta (3) \tau_2^{3/2}$$ 
and so~\cite{Green:1997me,Green:1998by}
\be \bar{D}_{-12} D_{11} f^{(11,-11)} (\tau, \bar\tau) = -\frac{25 \cdot 21}{16} f^{(11,-11)} (\tau, \bar\tau).\ee

Next consider the $\hat{G}^4 \lambda^{15} \g^\mu \psi^*_\mu$ interaction. We have looked at the part of the interaction
which involves $(\hat{G}^{\mu\nu\rho} \hat{G}_{\mu\nu\rho})^2$. However a similar analysis for the other space-time structures
shows that the modular forms multiplying them satisfy~\C{Lapeqn}\ as well with possibly different coefficients. However,
because these modular forms cannot receive perturbative contributions beyond genus two~\cite{Berkovits:2006vc}, 
and the genus one contribution vanishes~\cite{Green:1999pv}, we conclude that these modular forms satisfy the same Laplace
equation and contribute perturbatively only at genus zero and two. Thus they are all proportional to each other.
In fact, the complete spacetime structure can be deduced using topological string amplitudes and is 
proportional to~\cite{Sinha:2002zr}
\be \label{calcexp}
90 (\hat{G}^{\mu\nu\rho} \hat{G}_{\mu\nu\rho})^2 -15 \hat{G}_{\mu_1 \nu_1 \rho_1} \hat{G}^{\mu_1 \mu_3 \nu_2}
\hat{G}_{~\mu_3 \mu_4}^{\rho_1} \hat{G}^{~\mu_4 \mu_1}_{\nu_2} 
+\hat{G}_{\mu_1 \nu_1 \rho_1} \hat{G}^{\mu_1 \mu_2 \rho_1} \hat{G}_{\mu_3}^{~\nu_1\nu_2} \hat{G}_{~\mu_2 \nu_2}^{\mu_3}.\ee
Noting that the genus zero Einstein frame $D^4 \mathcal{R}^4$ interaction is proportional to $\zeta (5) \tau_2^{5/2}$, we see 
that 
$$f^{(13,-13)} \sim D_{12} \ldots D_0\zeta (5) \tau_2^{5/2} \sim \zeta (5) \tau_2^{5/2},$$ 
leading to~\cite{Sinha:2002zr} 
\be \bar{D}_{-14} D_{13}f^{(13,-13)} (\tau, \bar\tau) = -\frac{31 \cdot 23}{16} f^{(13,-13)} (\tau, \bar\tau).\ee
In fact, these modular forms for $k=0$ and $2$  are given by~\cite{Berkovits:1998ex}
\be f^{(q,-q)}_g (\tau, \bar\tau) = \sum_{(m,n) \neq (0,0)} 
\frac{\tau_2^{g+ 1/2}}{(m+ n \tau)^{g + q +1/2} (m + n \bar\tau)^{g-q + 1/2}},\ee
for $(g,q) =(1,11)$ and $(2,13)$ respectively. In this presentation, the modular forms satisfy the equation
\be 4\bar{D}_{-(q+1)}D_q f^{(q,-q)}_g (\tau, \bar\tau) 
= \Big(g + q + \frac{1}{2}\Big) \Big(g-q-\frac{1}{2}\Big) f^{(q,-q)}_g (\tau, \bar\tau) .\ee
We now turn to the contributions from the source terms.

\subsubsection{The source term contributions for $k=3$}

Let us first consider in some detail the $k=3$ case which is the first instance where the source 
term contributes. The various technical details and arguments are along the lines of the preceding discussion so we will only mention the
main results. The analysis involving $(S^{(6)}, S^{(0)})$ and $(\delta^{(6)}, \delta^{(0)})$ already appears in section~\ref{sugracontributions}. So we only need to consider the contributions involving $S^{(3)}$ and $\delta^{(3)}$.  

Among all the terms in $S^{(3)}$, there is only one term given by (dropping overall numerical factors)
\be L^{(3)}_1 = {\rm det} e f^{(3,-3)} (\tau, \bar\tau) \hat{G}^6 
(\bar\lambda^* \g^{\mu\nu\rho} \lambda )(\bar\lambda\g_{\mu\nu\rho} \lambda^*) \ee
which contributes non--trivially to the equations. The various other terms in $S^{(3)}$, as well as
their possible supervariations $\delta^{(3)}$, either do not give the required space-time structure, or are
inconsistent with closure of the superalgebra. Now $L^{(3)}_1$ gives no contribution under the supervariation $\delta^{(3)}$
into ${\rm det}e \hat{G}^6 \lambda^{16} (\bar\epsilon^* \g^\mu \psi_\mu^*)$. The contribution under $\delta^{(3)}$ into
${\rm det}e \hat{G}^6 \lambda^{16} (\bar\epsilon \lambda^*)$ is given by
\be \delta^{(3)} L^{(3)}_1 \sim q (\tau, \bar\tau) f^{(3,-3)} (\tau, \bar\tau) 
{\rm det}e \hat{G}^6 \lambda^{16} (\bar\epsilon \lambda^*), \ee
where
\be \delta^{(3)} \lambda^*_a \sim q (\tau, \bar\tau) (\lambda^{14})_{cd} 
(\g^{\mu\nu\rho} \g^0)_{dc} (\g_{\mu\nu\rho} \epsilon^*)_a .\ee
Closure of the supersymmetry algebra acting on $\lambda^*$ yields the equation of motion 
\be \g^0 \g^\mu D_\mu \lambda^* + \alpha'^3 (D_{11} q)  \hat{G}^{2k} \lambda^{15} + \ldots =0,\ee
thus leading to $D_{11} q \sim f^{(12,-12)}$. From the constraints for $k=0$, it therefore follows that
\be q (\tau, \bar\tau) \sim f^{(11,-11)} (\tau, \bar\tau) . \ee
Using this, as well as the previous constraints, we are finally led to the coupled differential equations
\bea \label{moreeqn}
&&D_{14} f^{(14, -14)} + \lambda_1 f^{(15, -15)} =0 , \non \\
&&\bar{D}_{-15} f^{(15, -15)} + \lambda_2 f^{(14, -14)} + \lambda_3 f^{(11,-11)} f^{(3,-3)} = 0.
\eea
Combining the equations, we see that coupling for the $\hat{G}^6 \lambda^{16}$ interaction satisfies 
\be \label{k3one}
D_{14} \bar{D}_{-15} f^{(15, -15)} = \alpha_1 f^{(15, -15)}  + \alpha_2 \left( f^{(12,-12)} f^{(3,-3)}
+ f^{(11,-11)} f^{(4,-4)} \right) ,\ee
while the coupling for the $\hat{G}^6 \lambda^{15} \g^\mu \psi_\mu^*$ and $\hat{G}^4 (\hat{G}^{\mu\nu\rho}
\hat{G}_{\mu\nu\rho}^*) \lambda^{16}$ interactions satisfy
\be \label{k3two}
\bar{D}_{-15} D_{14} f^{(14, -14)} = \alpha_1 f^{(14, -14)} + \alpha_3 f^{(11,-11)} f^{(3,-3)} ,\ee
where we have used $D_{11} f^{(11,-11)} = f^{(12,-12)}$, and $D_{3} f^{(3,-3)} = f^{(4,-4)}$. 
From~\C{k3one}\ and~\C{k3two}, we see that these couplings satisfy Poisson equations on moduli space sourced by interactions in $S^{(3)}$. In fact, the interaction involving $f^{(4,-4)}$ is given by
\be L^{(3)}_2 = {\rm det} e f^{(4,-4)} (\tau, \bar\tau) \hat{G}^8 .\ee 

Let us give a heuristic derivation of \C{moreeqn}, intuitively showing why this is the only possibility. Based
on considerations of $SL(2,\mathbb{Z})$ invariance and the fact that the set of interactions given in~\C{maincoup}\ involving at least fifteen dilatinos is special, we see that the most general system of equations that 
could have arisen from our analysis is (dropping various coefficients)
\be \label{intone}
D_{14} f^{(14, -14)} + f^{(15, -15)} + f^{(12,-12)} f^{(3,-3)} + f^{(11,-11)} f^{(4,-4)} \sim 0,\ee
\be \label{inttwo}
\bar{D}_{-15} f^{(15, -15)} +  f^{(14, -14)} + \hat{f}^{(14,-14)} + g_1 + p_1 + 
f^{(11,-11)} f^{(3,-3)} + f^{(12,-12)} f^{(2,-2)} \sim 0, \ee  
\be \label{intthree}
D_{14} g_1 \sim f^{(15,-15)}
, \qquad g_1 \sim \hat{f}^{(14,-14)}, \qquad g_1 + p_1 \sim f^{(14,-14)} .\ee  
From \C{intone} and \C{intthree}, we obtain 
\be D_{14} f^{(14, -14)} + f^{(12,-12)} f^{(3,-3)} + f^{(11,-11)} f^{(4,-4)} \sim D_{14} g_1,\ee
leading to 
\be 
g_1  \sim f^{(14,-14)} + f^{(11,-11)} f^{(3,-3)},\ee
from which one concludes that either both the terms involving 
$$f^{(12,-12)} f^{(3,-3)} \quad {\rm and} \quad f^{(11,-11)} f^{(4,-4)}$$ 
are present in \C{intone}, or both are absent. We therefore get the set of equations~\C{intone}\ and
\be \label{intfour}
\bar{D}_{-15} f^{(15, -15)} +  f^{(14, -14)} + f^{(11,-11)} f^{(3,-3)} + f^{(12,-12)} f^{(2,-2)} \sim 0.\ee
Acting with $D_{14}$ on \C{intfour}, we see that the last term involving $f^{(12,-12)} f^{(2,-2)}$ must be absent
because it gives a contribution 
$$D_{14} (f^{(12,-12)} f^{(2,-2)}) \sim f^{(13,-13)} f^{(2,-2)} + \ldots,$$ but there is no 
term in the $\mathcal{R}^4$ multiplet which has modular weight 13. As we shall see later, this argument does not 
generalize to higher $k$ because it is possible to have such a contribution from the $D^{2k'} 
\mathcal{R}^4$ multiplet for low enough 
$k'$. Finally, acting with $\bar{D}_{-15}$ on
\C{intone}, we get
\be \label{Smalleqn}
\bar{D}_{-15} D_{14} f^{(14, -14)}  + f^{(14, -14)} + f^{(11,-11)} f^{(3,-3)} + 
f^{(12,-12)} f^{(2,-2)} + f^{(10,-10)} f^{(4,-4)} \sim 0 .\ee
Now the last term in \C{Smalleqn} involving $f^{(10,-10)}$ originates from $\bar{D}_{-15}  (f^{(11,-11)} f^{(4,-4)})$
and involves the interaction
\be L^{(3)}_3 = {\rm det} e f^{(10,-10)} (\tau, \bar\tau) \hat{G}^2 \lambda^{12} .\ee
However, this interaction has only fourteen dilatinos and is not expected to mix with the special set of interactions 
we are considering. Thus the
terms involving both $f^{(12,-12)} f^{(3,-3)}$ and $f^{(11,-11)} f^{(4,-4)}$ are absent in \C{intone}, and we have obtained~\C{moreeqn},~\C{k3one}\ and~\C{k3two}. This generalizes to all $k$ modulo the preceding caveat.

As before,~\C{k3one}\ and~\C{k3two}\ have been obtained by considering a specific space-time structure for the 
relevant interactions. Considering other space-time structures, we see that they give the same equations though with possibly different coefficients. This phenomenon clearly occurs for higher $k$ as well as we shall see later. 
In fact, for $k \geq 4$, it is crucial that there is more than one modular form of a given modular weight
that arises this way.

The $k=3$ case is borderline and there might or might not be several modular forms for different space-time structures. This comes about because we know that the
modular forms cannot receive perturbative contributions beyond genus three~\cite{Berkovits:2006vc}. The genus three contribution of at least
one of the modular forms is expected to be 
non--zero since it should be related by supersymmetry to the genus three amplitude of the $D^6 \mathcal{R}^4$ 
interaction. This genus three contribution is non--vanishing and is given by one loop supergravity 
in eleven dimensions compactified on 
$T^2$~\cite{Green:1999pu,Green:1982sw,Russo:1997mk}\ after
using duality~\cite{Green:2006gt}. This computation actually gives the type IIA amplitude but the perturbative parts of the type IIA and 
type IIB contributions are the same for the $D^6 \mathcal{R}^4$ interaction~\cite{Berkovits:2006vc}.

 Now the source terms in~\C{k3one}\
and~\C{k3two}\ involving 
$$f^{(12,-12)},  \quad f^{(11,-11)}, \quad f^{(4,-4)}, \quad f^{(3,-3)}$$ 
contribute at genus zero, one and 
two only. The genus three contribution   
only enters the Laplace equation part of~\C{k3one}\ and~\C{k3two}, and so it is possible to 
have only one modular form which gets its perturbative contributions this way. However, it is also possible to have
more than one modular form where each form satisfies a separate Poisson equation with appropriate asymptotic behavior.

We know that the genus three contribution to $D^6 \mathcal{R}^4$ is non-vanishing~\cite{Green:2005ba}. So among the various contractions for $ {\hat G}^6$, there should be one coefficient function with a non-vanishing $3$-loop contribution. Let us consider that particular coefficient function. Because the genus three contribution involves only the Laplace equation part of \C{k3two}, taking
$f^{(14,-14)} \sim \tau_2^{-3}$ immediately leads to\footnote{This also has a solution $\tau_2^4$ which is inconsistent
with string perturbation theory.} 
\be \label{oneval}
\alpha_1 = -\frac{99}{2}. \ee

We now fix $\alpha_3$ using the expression for the genus zero amplitude in \C{treeamp}. The $f^{(14,-14)}$ amplitude
is related by supersymmetry to the $D^6 \mathcal{R}^4$ amplitude and so we take 
$$f^{(14,-14)} = \zeta (3)^2 \tau_2^3 +\ldots,$$
up to an overall irrelevant constant. We need the perturbative parts of the weak coupling expansions 
for $f^{(11,-11)}$ and $f^{(3,-3)}$ 
which are obtained by using 
\be \Gamma \Big( \frac{9}{2} \Big) f^{(3,-3)} = 4\sqrt{\pi} D_2 D_1 D_0 E_{3/2} , \quad \Gamma 
\Big( \frac{25}{2} \Big) f^{(11,-11)}
= 2^{10} \sqrt{\pi} D_{10} \ldots D_0 E_{3/2},\ee
leading to
\bea f^{(3,-3)} (\tau, \bar\tau) = 2 \zeta (3) \tau_2^{3/2} -\frac{4}{35} \zeta (2)
\tau_2^{-1/2} + \ldots, \non \\ 
f^{(11,-11)} (\tau, \bar\tau) = 2 \zeta (3) \tau_2^{3/2} -\frac{4}{483} \zeta (2) \tau_2^{-1/2} 
+ \ldots, \eea
after using~\C{expSL2}. 

Equating terms of $O(\tau_2^3)$ on both sides of \C{k3two} and using \C{oneval}, we find that
\be \alpha_3 = -\frac{3}{8}.\ee
Similarly after equating terms of $O(\tau_2)$ and $O(\tau_2^{-1})$, we can calculate the genus 
one and genus two contributions to $f^{(14,-14)}$.
The calculation of the genus three contribution is 
involved. It should be possibly by generalizing a similar calculation in~\cite{Green:2005ba}. 
Then one can also determine the non--perturbative 
contributions to $f^{(14,-14)}$ which involves contributions from single D--instantons, and D-instanton/D-anti-instanton pairs. The first equation of~\C{moreeqn}\ then determines $f^{(15,-15)}$. 

We can say something additional about the coefficient functions for other space-time contractions. Suppose we consider a different coefficient function $ {\hat f}^{(14,-14)}$. Then it must satisfy
\be \label{neweqn}
\bar{D}_{-15} D_{14} {\hat f}^{(14,-14)} = \s_1 {\hat f}^{(14,-14)} + \s_2 {f}^{(11,-11)} {f}^{(3,-3)} .\ee
Again using  Berkovits' theorem~\cite{Berkovits:2006vc}, we find that the perturbative part of ${\hat f}^{(14,-14)}$ is given by
\be {\hat f}^{(14,-14)} \sim \tau_2^3 + \tau_2 + \tau_2^{-1} + \tau_2^{-3} + \ldots.\ee
If the $3$-loop contribution to ${\hat f}^{(14,-14)}$ is non-vanishing then it follows that 
\be \s_1 = -\frac{99}{2}\ee
and therefore $ {\hat f}^{(14,-14)}$ is proportional to  ${f}^{(14,-14)}$. If the $3$-loop contribution does vanish then the coefficient function can be linearly independent.

Employing supersymmetry, we see that the coefficient of the $D^6 \mathcal{R}^4$ interaction 
can also be a sum of modular forms $f^{(0,0)\{i\}}$, only one of which must
receive a perturbative contribution at genus three (call it $f^{(0,0)\{0\}}$). 
Furthermore, each of these modular forms must satisfy     
\be \label{calcmoreyet}
4 \tau_2^2 \frac{\p^2}{\p \tau \p \bar\tau} f^{(0,0)\{i\}} (\tau,\bar\tau) = \lambda_{1i} f^{(0,0)\{i\}} 
(\tau,\bar\tau) + \lambda_{2i} E_{3/2} (\tau,\bar\tau) E_{3/2} (\tau,\bar\tau) . \ee
Based on our discussion above, the source terms in \C{calcmoreyet} are uniquely fixed by the modular properties of 
\C{calcmoreyet}. Since $f^{(0,0)\{i\}}$ has weight $(0,0)$, the sources must also involve weight $(0,0)$ forms because there are no modular
forms of weight $(-q,q)$ to pair with forms of weight $(q,-q)$ for any non--zero $q$. This follows because both coefficient functions multiply ``chiral''
couplings where by ``chiral'' (``antichiral'') we mean those couplings which have weight $(q,-q)$ for positive (negative)
$q$, or equivalently positive (negative) $U(1)$ charge. For example, the $\lambda^{16}$ coupling $f^{(12,-12)}$ is chiral, whereas the $(\lambda^*)^{16}$ coupling $f^{(-12,12)}$
is not. Note that none of these couplings are holomorphic with respect to $\tau$. 

Therefore $f^{(0,0)\{0\}}$ satisfies
\be \label{calcmoreyet2}
4 \tau_2^2 \frac{\p^2}{\p \tau \p \bar\tau} f^{(0,0)\{0\}} (\tau,\bar\tau) = 12 f^{(0,0)\{0\}} 
(\tau,\bar\tau) + \hat\lambda E_{3/2} (\tau,\bar\tau) E_{3/2} (\tau,\bar\tau) . \ee
Extracting the tree level contribution, we see that $f^{(0,0)\{0\}} = 4 \zeta (3)^2 \tau_2^3$ which implies that $\hat\lambda =-6$. 
One can immediately compute the remaining perturbative contributions to $f^{(0,0)\{0\}}(\tau, \bar\tau)$. These contributions have been computed in~\cite{Green:2005ba}. The numerical coefficients obtained from~\C{calcmoreyet2} for 
the genus one amplitude matches the string theory computation while the genus three contribution matches the supergravity computation exactly; the genus two amplitude is not known by either a direct string or supergravity calculation. Thus if there 
are any other modular forms $f^{(0,0)\{i\}}$, they can only receive a perturbative contribution at genus two which is inconsistent 
with~\C{calcmoreyet}; hence they vanish.

\subsubsection{The source term contributions for $k>3$}
 
We now derive the structure of the equations which determine the couplings for general $k$. Some of the analysis is 
similar to the $k=3$ case so we will be brief. We already know the contributions
from $S^{(k+3)}$ and $S^{(0)}$ determined in section~\ref{sugracontributions}, so we focus only on the source terms. Consider terms in the action $S^{(k-k')}$, where
$k' < k$. There are two kinds of interactions in $S^{(k-k')}$ which are relevant for us.

The first kind of interaction involves the terms 
\bea L^{(k-k')}_1 = {\rm det} e f^{(k-k',k' -k)} (\tau, \bar\tau) \hat{G}^{2(k-k')} 
(\bar\lambda^* \g^{\mu\nu\rho} \lambda )(\bar\lambda\g_{\mu\nu\rho} \lambda^*) , \non \\
L^{(k-k')}_2 = {\rm det} e {\hat f}^{(k-k',k' -k)} (\tau, \bar\tau) \hat{G}^{2(k-k')} 
(\bar\psi^*_{[\mu} \g_\nu \psi_{\rho]}) (\bar\lambda \g^{[\mu\nu} \psi^{\rho]}) . \eea
Under the supervariation $\delta^{(k' +3)}$, they do not vary into ${\rm det}e \hat{G}^{2k} 
(\bar\epsilon^* \g^\mu \psi_\mu^*) \lambda^{16}$. However, they vary into
\be \delta^{(k' +3)} (L^{(k-k')}_1 + L^{(k-k')}_2) \sim \left\{ t_1 f^{(k-k',k' -k)} 
+ t_2 {\hat f}^{(k-k',k' -k)}   \right\}
{\rm det}e \hat{G}^{2k} \lambda^{16} (\bar\epsilon \lambda^*) ,\ee
where
\bea \delta^{(k'+3)} \lambda^*_a &=& t_1 (\tau, \bar\tau) \hat{G}^{2k'}(\lambda^{14})_{cd} 
(\g^{\mu\nu\rho} \g^0)_{dc} (\g_{\mu\nu\rho} \epsilon^*)_a, \non \\
\delta^{(k'+3)} \psi_\mu^a &=& t_2 (\tau, \bar\tau) \hat{G}^{2(k'-1)} \lambda^{16}  
(\g^{\s_1 \s_2 \s_3} \g_\mu \epsilon^*)^a \hat{G}_{\s_1 \s_2 \s_3}. \eea 
The other kind of interaction involves
\be \label{onemore}
L^{(k-k')}_3 = {\rm det} e f^{(k-k' -1,k' -k +1)} (\tau, \bar\tau) \hat{G}^{2(k-k' -1)}
(\hat{G}^{\mu\nu\rho} \hat{G}^*_{\mu\nu\rho})^2. \ee
Under the supervariation $\delta^{(k' +3)}$, this also does not vary into ${\rm det}e \hat{G}^{2k}
(\bar\epsilon^* \g^\mu \psi_\mu^*) \lambda^{16}$ but it does vary into
\be \delta^{(k' +3)} L^{(k-k')}_3 \sim f^{(k-k' -1,k' -k +1)} (\tau, \bar\tau) t_3 (\tau, \bar\tau) 
{\rm det}e \hat{G}^{2k} \lambda^{16} (\bar\epsilon \lambda^*) ,\ee
where
\be \delta^{(k'+3)} \lambda^*_a =t_3 (\tau, \bar\tau) \hat{G}^{2(k' -1)} \lambda^{16} 
\hat{G}_{\mu\nu\rho}  (\g^{\mu\nu\rho} \epsilon^*)_a , \ee
after using the Fierz identity. From \C{onemore}, it follows that this contribution exists only for $k' <k -1$, 
and is absent for the $k=3$ case. 

From closure of the superalgebra, we obtain\footnote{In calculating the closure condition involving 
$t_3$, we use $\delta^{(0)} \lambda \sim \epsilon \hat{G}$.}
\bea & D_{k' + 11} t_1 \sim f^{(12 + k',-12-k')} ,&  \qquad t_1 \sim \hat{f}^{(11 + k', -11 - k')}, \non \\
& t_1 + t_2 \sim f^{(11 + k', -11 - k')}, & \qquad  t_3 \sim f^{(12+k', -12 -k')}\eea
which leads to
\be t_1 \sim t_2 \sim f^{(11 + k', -11 -k')} \sim \hat{f}^{(11 + k', -11 -k')}.\ee
Taking into account these source term contributions for all values of $k'$,
we find the equations (again ignoring the numerical coefficients)
\bea \label{decouple}
&&D_{11 + k} f^{(11 + k, -11 -k)} + f^{(12 + k, -12 -k)} =0 , \non \\
&&\bar{D}_{-(12 + k)} f^{(12 + k, -12 -k)} + f^{(11 + k, -11 -k)} \non \\ &&+ \sum_{k'} 
\Big( f^{(11+ k',-11-k')} f^{(k-k',k'-k)} + f^{(12+ k',-12-k')} f^{(k-k'-1,k'-k+1)} \Big) = 0.
\eea
Note that the structure of the second equation in \C{decouple} is strikingly similar to the 
holomorphic anomaly equation satisfied by certain protected interactions 
in the effective action of $N=2$ string theory~\cite{Bershadsky:1993cx}. This is unlikely to be
an accident!

The coupling for the $\hat{G}^{2k} \lambda^{16}$ interaction therefore satisfies the equation
\bea \label{karbone}
&&D_{11 +k} \bar{D}_{-(12 + k)} f^{(12 +k, -12-k)} = a_k f^{(12+k, -12-k)}  \non \\
&&+D_{11 +k} \sum_{k'} 
\Big( b_{kk'} f^{(11+ k',-11-k')} f^{(k-k',k'-k)} + c_{kk'} f^{(12+ k',-12-k')} f^{(k-k'-1,k'-k+1)} \Big),\qquad\eea
while the coupling for the $\hat{G}^{2k} \lambda^{15} \g^\mu \psi_\mu^*$ 
and $\hat{G}^{2(k-1)} (\hat{G}^{\mu\nu\rho} \hat{G}_{\mu\nu\rho}^*) \lambda^{16}$ interactions satisfy
\bea \label{karbtwo}
&&\bar{D}_{-(12+k)} D_{11 +k} f^{(11+k, -11-k)} =  a_k f^{(11+k , -11-k)} \non \\
&&+ \sum_{k'} \Big( d_{kk'} f^{(11+ k', -11 -k')} f^{(k-k', k'-k)} +
e_{kk'} f^{(12+ k',-12-k')} f^{(k-k'-1,k'-k+1)}\Big),\eea
where $a_k, b_{kk'},c_{kk'}, d_{kk'}$, and $e_{kk'}$ are undetermined coefficients. 
We should note that the source terms in~\C{karbone}\ involve $D_{11+k}$ acting on a product of two modular forms. This can give rise to sources that are cubic in modular forms.\footnote{We would like to thank M.~B.~Green for discussions explaining this point.} 
These cubic source terms first appear for couplings of order $D^{12} {\mathcal R}^4$ since the source terms involve products of a covariant derivative acting on a coefficient function from order $D^{6} {\mathcal R}^4$ multiplied with a coefficient function from order ${\mathcal R}^4$~\cite{Basu:2006cs}; see also~\cite{Chalmers:2006aj}\ for related comments. 

As before, in deducing equations \C{karbone} and \C{karbtwo}, we have focused on the modular forms associated with 
interactions where the space-time structure involves $\hat{G}^2 = \hat{G}^{\mu\nu\rho} \hat{G}_{\mu\nu\rho}$. 
Modular forms of the same modular weights associated with other space-time structures will satisfy the same equations 
but with possibly different coefficients. So we should label these other coefficient functions with an extra index, which we will ignore for the sake of simplicity. 

\subsection{Sufficiently large $k$}

The discussion above is strictly valid for sufficiently small $k$, since we cannot write $\hat{G}^{2k} 
\sim (\psi \psi)^{2k}$ for arbitrarily large $k$ because of the exclusion principle. Noting 
that 
$$\bar\psi^*_\mu \g_\nu \psi_\rho = \psi_\mu^a (\g^0 \g_\nu)_{ab} \psi_\rho^b$$ 
where $(\g^0 \g_\nu)_{ab}$ 
is a symmetric matrix, and that $\psi_\mu^a$ allows for 160 indices to be assembled, we see that for $k > 40$, we 
have to take
\be \hat{G}^{2k} = (G_{\s_1 \s_2 \s_3} G^{\s_1 \s_2 \s_3})^{(k-40)} (-36 \bar\psi^*_{[\mu} \g_\nu \psi_{\rho]} 
\bar\psi^{*[\mu} \g^\nu \psi^{\rho]} )^{40} + \ldots \equiv  G^{2(k-40)}(-6i \psi \psi)^{80}
+ \ldots.\ee
Thus $N=40$ and we can absorb 80 powers of the derivative in the fermions and the remaining powers in $G$. However, the explicit
value of $k$ where this transition occurs is not really needed in our analysis.   

We again consider the interactions
\bea 
L^{(k+3)}_1 &=& {\rm det} e  f^{(12+k,-12-k)} (\tau,\bar\tau) G^{2(k-40)} (-6i \psi \psi)^{80} \lambda^{16},\non \\
L^{(k+3)}_2 &=& {\rm det} e f^{(11+k,-11-k)} (\tau,\bar\tau) G^{2(k-40)}  (-6i \psi \psi)^{80} \lambda^{15} \g^\mu \psi_\mu^* ,
\non \\ L^{(k+3)}_3 &=& {\rm det} e \hat{f}^{(11+k,-11-k)} (\tau,\bar\tau) G^{2(k-40)}  (-6i \psi \psi)^{78}
(\hat{G}^{\mu\nu\rho} \hat{G}_{\mu\nu\rho}^*)\lambda^{16}. \eea
The analysis is very similar to the $k<N$ case and leads to the same conclusions. In various places, there are some modifications needed with
$$ 
(-6i \psi \psi)^{2k} \rightarrow G^{2(k-40)} (-6i \psi\psi)^{80}, \qquad (-6i \psi \psi)^{2(k-1)} \rightarrow G^{2(k-40)}  (-6i \psi \psi)^{78}.
$$
The only issue
is to explicitly see the appearance of modular covariant derivatives with the correct modular weights in
the supervariations and the closure of the superalgebra. This happens
by adding  a compensating $U(1)$ gauge transformation in the supervariation of $G_{\mu\nu\rho}$ given 
by
\be \label{compU1}
\delta^{(0)} G_{\mu\nu\rho} = \frac{i}{2} (\bar\epsilon \lambda^* - \bar\epsilon^* \lambda)G_{\mu\nu\rho}, \ee
to that given in~\C{impdef}\ and~\C{addtodef}. This is fixed by the fact that $G_{\mu\nu\rho}$ has $U(1)$ charge 1. The resulting equations for the couplings take the same form given in~\C{karbone}\ and \C{karbtwo}.

\section{Some Consequences of the Poisson Equations}
 \label{implications}
 
We now explore some consequences of the Poisson equations~\C{karbone}\ and~\C{karbtwo}. We will argue that these 
protected interactions satisfy a perturbative non--renormalization theorem.
We also demonstrate various
qualitative features of specific protected interactions based on constraints of unitarity and known perturbative 
amplitudes.  

\subsection{A perturbative non--renormalization theorem}

From the structure of either~\C{karbone}\ or~\C{karbtwo}, we want to argue that our special couplings can receive only a finite number of 
perturbative contributions. This follows from noting that for low values of $k$, the couplings have this property as we have explicitly seen. We can then apply induction to argue the same result for all $k$ because the source terms themselves at each step only involve a finite number of perturbative contributions.

The remaining issue is to constrain perturbative contributions for the terms multiplying the $a_k$ coefficients which are present in the absence of the source terms. So we consider the source-free Laplace equation. Solutions to this equation receive at most two perturbative contributions which completes the argument. 

As we have discussed earlier, it seems quite plausible that this special renormalization property will extend to all terms in the supermultiplet of couplings related by supersymmetry. So we might reasonably expect that the $D^{2k} \mathcal{R}^4$ coupling has this property. Regardless, we can conclude that there are an infinite number of protected interactions in 
type IIB string theory. Each interaction receives only a finite number of perturbative 
contributions together with a collection of non--perturbative contributions.

\subsection{The $k  = 4$ case and aspects of the $D^8 \mathcal{R}^4$ interaction}

Let us analyse the $k=4$ case in more detail. This is the first case where unitarity constraints require a
new type
of perturbative contribution to modular forms which multiply interactions that are not vanishing on--shell. The source terms
for the $D^8 \mathcal{R}^4$ interaction are given by a particular 
interaction in $S^{(4)}$ (related by supersymmetry to $\hat{G}^2 \lambda^{16}$) 
that vanishes on--shell, but is needed
on the basis of unitarity constraints~\cite{Green:2008bf}. To see this, as well as to understand 
some aspects of the prototype $D^8 \mathcal{R}^4$ interaction, let us briefly 
discuss non--local terms in the effective action.   

Since we are looking at the 1PI effective action, we allow massless modes to propagate in loop amplitudes. This
leads to terms in the effective action which are non--analytic in the external momenta and are therefore non--local.
This problem can be avoided by looking at the Wilsonian rather than 1PI effective action but at the cost of sacrificing duality invariance.

The behavior of this non--analyticity is dimension-dependent. In ten dimensions, it is logarithmic. In a string-frame scattering computation with string metric, $ g_{\mu\nu} $, there are terms involving
\be
{\rm ln}(s) ={\rm ln} ( g^{\mu\nu} k_\mu k_\nu) = {\rm ln} ( \sqrt{\tau_2} \, {\hat g}^{\mu\nu} k_\mu k_\nu) =  {\rm ln} ( \sqrt{\tau_2} \, {\hat s}) = {1\over 2} {\rm ln}(\tau_2) + {\rm ln}( {\hat s})
\ee
where ${\hat g}_{\mu\nu}$ denotes the Einstein frame metric. So we can attribute  any  $ {\rm ln} (\tau_2)$ terms in the coefficient functions of local couplings in Einstein frame to non-local interactions in string frame. 
These non-local interactions can therefore contribute to the modular forms for the $D^{2k} \mathcal{R}^4$ and related interactions, and must be considered in our analysis. 

Some of these non--local terms in the effective action have been analyzed based on 
unitarity~\cite{Green:2006gt}, and the first few are schematically given in the string frame by
\be  \label{nlt}
\alpha'^4 \Big( s{\rm ln} (-\alpha' s) + \alpha'^3 \tau_2^{-3/2} E_{3/2} s^4 {\rm ln} (-\alpha' s)
+ \alpha'^4 \tau_2^{-2} s^5 {\rm ln}^2 (-\alpha' s) + \ldots \Big) \mathcal{R}^4,\ee
where we have dropped the additional terms needed to symmetrize in $s,t$, and $u$ for brevity. 
The $O(\alpha'^4)$ contribution in~\C{nlt}\ is at genus one, the $O(\alpha'^7)$ contribution is at genus one and two, while the
$O(\alpha'^8)$ contribution is at genus two only. In Einstein frame, these non-local terms make a contribution from
\be \label{nlt2}
\alpha'^4 \Big( {\rm ln} \tau_2 {\hat s} + \alpha'^3  E_{3/2} {\rm ln} \tau_2 {\hat s}^4
+ \alpha'^4 ({\rm ln} \tau_2 )^2 {\hat s}^5 + \ldots \Big) \mathcal{\hat R}^4\ee
to the local terms in the effective action. 
So we see that the modular form for the $D^8 \mathcal{R}^4$ interaction  receives contributions 
logarithmic in $\tau_2$. It is reasonable to expect that there might be similar logarithmic terms in  the 
$\hat{G}^8 \lambda^{15} \g^\mu \psi_\mu^*$ interaction.

Note that the first term in \C{nlt2} vanishes on--shell using $s+t+u=0$; however, we need to consider
its effect as a source term for the higher derivative interactions~\cite{Green:2008bf}. In fact, integrating by parts, we see that this term does survive in the effective action for non--constant $\tau$ at order $D^{2} \mathcal{R}^4$. We denote the
complete modular form for  $D^{2} \mathcal{R}^4$ by $Z(\tau, \bar\tau)$, remembering that it receives a perturbative contribution only at
genus one proportional to ${\rm ln} (\tau_2)$. By acting on this modular form with a suitable number of modular 
covariant derivatives, we get source terms for the Poisson equations satisfied by 
the various protected interactions.  

Returning to the interactions in~\C{karbtwo}\footnote{There is a similar analysis for
\C{karbone}.}, it follows from~\cite{Berkovits:2006vc}\ both that there are no perturbative
contributions beyond genus four, and that the perturbative contributions are the same in type IIA 
and type IIB. From~\cite{Green:2006gt}, we know that the  
genus four amplitude for the $D^8 \mathcal{R}^4$ interaction in type IIA string theory 
is non--vanishing; consequently, we expect the genus four $\hat{G}^8
\lambda^{15} \g^\mu \psi_\mu^*$ amplitude is non--vanishing as well by supersymmetry. The genus four contribution to $D^8 \mathcal{R}^4$
is completely determined by the one loop 
four graviton amplitude in $d=11$ supergravity compactified on $T^2$ together with duality~\cite{Green:1999pu,Green:1982sw,Russo:1997mk}. 

Because the genus zero and genus
four amplitudes are both non--vanishing, and the source terms in \C{karbtwo} come from the products of modular forms
in the $\mathcal{R}^4$ 
and $D^2 \mathcal{R}^4$ supermultiplets which contribute only at genus one and two, there cannot be a single coefficient function satisfying a single Poisson equation: there must be at least distinct two modular 
forms! 

The different spacetime structures involving $\hat{G}^8 \lambda^{15} \g^\mu \psi_\mu^*$ 
must give at least two independent modular forms of the same modular weight. One of them has a genus zero contribution
$\sim \tau_2^{7/2}$ (but no genus four contribution) and satisfies
\be \label{8eqn1}
\bar{D}_{-16} D_{15} {\hat f}^{(15,-15)}_1 (\tau, \bar\tau) = -\frac{37 \cdot 25}{16} {\hat f}^{(15,-15)}_1 (\tau, \bar\tau)
+ {\rm sources},\ee
while the other has a genus four contribution $\sim \tau_2^{-9/2}$ (but no genus zero or genus three contributions) and satisfies
\be \label{8eqn2}
\bar{D}_{-16} D_{15} {\hat f}^{(15,-15)}_2 (\tau, \bar\tau) = -\frac{21^2}{16} {\hat f}^{(15,-15)}_2 (\tau, \bar\tau)
+ {\rm sources}.\ee
It would be interesting to understand whether a logarithmic $\tau_2$-dependence appears in the $\hat{G}^8 \lambda^{15} \g^\mu \psi_\mu^*$ or $\hat{G}^8 \lambda^{16}$ interactions. Our analysis implies that there can be
a logarithmic dependence only if the sources from $D^2 \mathcal{R}^4$ contain a logarithmic dependence. In principle, this can be determined by direct computation of all the sources at order $D^2 \mathcal{R}^4$. 

There are alternative approaches that involve direct computation. The first is to perform an explicit calculation of the genus one amplitude in ten dimensions. The second involves compactifying on a circle and studying the
ten-dimensional limit to analyze the contribution to the threshold corrections that come from the Kaluza-Klein modes. Both these calculations are technically involved. 

What we can conclude is that the various space-time structures in the ${\rm det}e \hat{G}^8 \lambda^{16}$ interaction
must yield at least two independent 
modular forms, even though each separate space-time structure gives rise to only one modular form.
That the $D^8 \mathcal{R}^4$ interaction has a unique space-time structure can be seen from~\C{treeamp}.  
The discussion above shows that the coefficient function $\hat{f}^{(0,0)}_{D^8 \mathcal{R}^4}$, which multiplies this $D^8 \mathcal{R}^4$ interaction, 
must split into at least two modular forms
\be \hat{f}^{(0,0)}_{D^8 \mathcal{R}^4} = \hat{f}^{(0,0)}_1 + \hat{f}^{(0,0)}_2 + \ldots.\ee      
The modular forms, $\hat{f}^{(0,0)}_1$ and $\hat{f}^{(0,0)}_2$, receive perturbative contributions at genus zero (but not at genus four) and genus four (but not at genus zero or three) respectively. They
satisfy the equations
\bea \label{incsource} 
4 \tau_2^2 \frac{\p^2}{\p \tau \p \bar\tau} \hat{f}^{(0,0)}_1 (\tau, \bar\tau)= \frac{35}{4} \hat{f}^{(0,0)}_1 
(\tau, \bar\tau)+ \nu_1
E_{3/2} (\tau, \bar\tau) Z (\tau, \bar\tau) , \non \\ 
4 \tau_2^2 \frac{\p^2}{\p \tau \p \bar\tau} \hat{f}^{(0,0)}_2 (\tau, \bar\tau)
= \frac{99}{4} \hat{f}^{(0,0)}_2 (\tau, \bar\tau)
+ \nu_2 E_{3/2} (\tau, \bar\tau) Z (\tau, \bar\tau).\eea    
From \C{treeamp}, ignoring overall coefficients, we take
\be \hat{f}^{(0,0)}_1 = \zeta (7)  \tau_2^{7/2} + \ldots. \ee
Substituting into~\C{incsource}\ gives
\bea \label{fixval}
\hat{f}^{(0,0)}_1 &=&  \zeta (7) \tau_2^{7/2} - \frac{\nu_1}{16} \zeta (3) \tau_2^{3/2} (1 + 4 ~{\rm ln} \tau_2)
+ \frac{\nu_1 \pi^2}{48} \tau_2^{-1/2} (1-4 ~{\rm ln} \tau_2) 
+a_3 \tau_2^{-5/2} +\ldots, 
\non \\ \hat{f}^{(0,0)}_2 &=&  
- \frac{\nu_2}{144} \zeta (3) \tau_2^{3/2} (1 + 12 ~{\rm ln} \tau_2)
+ \frac{\nu_2 \pi^2}{432} \tau_2^{-1/2} (1-12 ~{\rm ln} \tau_2) 
+a_4 \tau_2^{-9/2}+\ldots,\eea
using 
\be Z(\tau, \bar\tau) = {\rm ln} \tau_2 + \ldots. \ee
In \C{fixval}, $a_3$ and $a_4$ are the genus three and genus four contributions, respectively, while the dots represent contributions from D--instantons.

Note that \C{fixval} has ${\rm ln} \tau_2$ contributions at genus one and two, which is
consistent with the structure in \C{nlt2}. Also the $\tau_2^{3/2}$ part of the genus one amplitude for the 
$D^8 \mathcal{R}^4$ interaction vanishes~\cite{Green:2008uj}, and so the sum of the contributions to $\tau_2^{3/2}$  from all the modular forms must vanish.

There is an interesting observation that follows if we assume that there are precisely two modular forms for $D^8 \mathcal{R}^4$. This is the minimal possible number. In this case, $\nu_1$ and $\nu_2$ are related by the condition
\be
{\nu_2} = - 9 \nu_1
\ee 
which ensures that the $\tau_2^{3/2}$ contribution vanishes. However this also implies that the $\tau_2^{-1/2}$ contribution vanishes which leaves only the non-analytic contribution at genus two as well as genus one. The genus two contribution is currently unknown. It would be interesting to see if this is indeed the case.

\subsection{The $k \geq5$ case and aspects of the $D^{10} \mathcal{R}^4$ interaction}

As a final application, let us consider the $k=5$ case and impose the constraint that the modular forms 
cannot receive perturbative contributions beyond 
genus five~\cite{Berkovits:2006vc}. Also the genus five amplitude is non--vanishing in type 
IIA string theory, as can be seen from the one loop four graviton scattering amplitude in $d=11$ supergravity 
on $T^2$~\cite{Green:1999pu}. So it is natural to expect that the genus five type IIB amplitude is non--vanishing 
as well.\footnote{The proof by Berkovits~\cite{Berkovits:2006vc}\ demonstrating perturbative equality for type IIA and type IIB
stops at $k=4$. It can probably be extended to higher values of $k$ along the lines of~\cite{Berkovits:2006vi}.} 
In fact, if one considers the supermultiplet generated by the
part of the four graviton scattering amplitude 
involving contributions only from the even--even spin structures (which involve the space-time
structure $t_8 t_8 R^4$), the perturbative equality follows trivially for all $k$. The difference arises 
for the odd--odd spin structure contributions (which involve the space-time structure $\epsilon_{10}
\epsilon_{10} R^4$). The tree level amplitude in~\C{treeamp}\ involves the even--even spin structure contributions, and we focus only on that part in the discussion below.     

Consider~\C{karbone}\ where the source terms arise in two different ways: (i) they either involve the
products of modular forms in the $\mathcal{R}^4$ and $D^4 \mathcal{R}^4$ supermultiplets, or (ii) the squares of modular 
forms in the $D^2 \mathcal{R}^4$ supermultiplet. In either case, the source terms receive perturbative contributions 
up to genus three only. 
Assuming that the genus four contribution is non--vanishing, it follows that the genus four and five
contributions must both be given by the Laplace equation part of~\C{karbone}, which is not possible. 

It immediately
follows that there should be at least two modular forms ${\tilde f}^{(17,-17)}_1$ and ${\tilde f}^{(17,-17)}_2$ which receive 
perturbative contributions at genus four $\sim \tau_2^{-4}$
and genus five $\sim \tau_2^{-6}$ 
respectively. Thus they satisfy the equations
\bea D_{16} \bar{D}_{-17} {\tilde f}^{(17, -17)}_1 = -63 {\tilde f}^{(17, -17)}_1 + {\rm sources},\non \\
D_{16} \bar{D}_{-17} {\tilde f}^{(17, -17)}_2 = -\frac{115}{2} {\tilde f}^{(17, -17)}_2 + {\rm sources}.\eea
Note that ${\tilde f}^{(17,-17)}_1$ (${\tilde f}^{(17,-17)}_2$) does not contain a perturbative contribution at genus five (four). 
There can be more modular forms ${\tilde f}^{(17,-17)}_i$, some of which receive perturbative contributions up to 
genus three, while others receive perturbative contributions at genus four or five (but not both). 

As before, different space-time structures in ${\rm det}e \hat{G}^{10} \lambda^{16}$ must yield at least two independent 
modular forms, though each separate space-time structure gives rise to only one modular form.  
This phenomenon continues for higher $k$ as well, unless some perturbative contributions vanish for a specific 
value of $k$. 

What does this imply for the $D^{10} \mathcal{R}^4$ interaction? From \C{treeamp}, we see
that there is again a unique space-time structure, and so the discussion above shows that the modular form  
$f^{(0,0)}_{D^{10} \mathcal{R}^4}$ multiplying this interaction must split into at least two modular forms
\be f^{(0,0)}_{D^{10} \mathcal{R}^4} \sim {\tilde f}^{(0,0)}_1 + {\tilde f}^{(0,0)}_2 + \ldots,\ee      
where ${\tilde f}^{(0,0)}_1$ ($ {\tilde f}^{(0,0)}_2$) receives perturbative contributions at genus four (five), but not at genus five (four). 
In fact, they must satisfy
\bea \label{calcmore} 4 \tau_2^2 \frac{\p^2}{\p \tau \p \bar\tau} {\tilde f}^{(0,0)}_1 (\tau,\bar\tau) = 20 {\tilde f}^{(0,0)}_1 
(\tau,\bar\tau) + \mu_1 E_{3/2} (\tau,\bar\tau) E_{5/2} (\tau,\bar\tau) + \lambda_1 Z(\tau,\bar\tau)^2,\non \\
4 \tau_2^2 \frac{\p^2}{\p \tau \p \bar\tau} {\tilde f}^{(0,0)}_2 (\tau,\bar\tau)= 42 {\tilde f}^{(0,0)}_2 
(\tau,\bar\tau) + \mu_2 E_{3/2} (\tau,\bar\tau) E_{5/2}(\tau,\bar\tau)+ \lambda_2 Z(\tau,\bar\tau)^2. \eea 
The source terms in~\C{calcmore}\ are determined from the constraints that at this order they be (a) quadratic in lower coefficient functions with non-negative $U(1)$ charge (b) of modular weight zero. These are sufficient conditions to determine the sources at this order in the momentum expansion. 

Let us analyze \C{calcmore} in some detail. From \C{treeamp}, ignoring overall coefficients, we take
\be {\tilde f}^{(0,0)}_i = \zeta (3) \zeta (5) \tau_2^4 + \ldots, \ee
for $i=1,2$. 
This immediately leads to 
\bea \mu_1 = -2, \quad \mu_2 = -\frac{15}{2},~~~~~~~~~~~~~~~~~~~~~~~~~~~~~~~~~~~~ \non \\ 
{\tilde f}^{(0,0)}_1 = \zeta (3) \zeta (5) \tau_2^4 + \frac{8}{9} \zeta (2) \zeta (5) \tau_2^2 
+ \Big[ \frac{8}{15} \zeta (3) \zeta (4) - \frac{\lambda_1}{20} \Big( \frac{21}{200} - \frac{1}{10}
~{\rm ln} \tau_2 +({\rm ln} \tau_2)^2 \Big) \Big] \non \\
+ \frac{32}{21} \zeta (2) \zeta (4) \tau_2^{-2}  + a_4 \tau_2^{-4} + \ldots, \quad 
\non \\ {\tilde f}^{(0,0)}_2 =  \zeta (3) \zeta (5) \tau_2^4 + \frac{3}{2} \zeta (2) \zeta (5) \tau_2^2 
+ \Big[ \frac{20}{21} \zeta (3) \zeta (4) - \frac{\lambda_2}{42} \Big( \frac{43}{882} - \frac{1}{21}
~{\rm ln} \tau_2 +({\rm ln} \tau_2)^2 \Big)\Big] \non \\ 
+ \frac{20}{9} \zeta (2) \zeta (4) \tau_2^{-2} + a_5 \tau_2^{-6} + \ldots,\quad\eea 
where $a_4$ and $a_5$ are the genus four and five contributions, respectively. The dots represent 
contributions from D--instantons. 

From~\C{nlt2}, we see that the total contribution to the genus two amplitude proportional to ${\rm ln} \tau_2$ 
vanishes so the sum of all such contributions from the various modular forms must vanish.
The genus one contribution proportional to $\zeta (2) \zeta (5)$ 
is consistent with a direct string one loop calculation~\cite{Green:2008uj}. 
Though we do not have sufficient perturbative data to fix
$a_4$ and $a_5$, we can keep only the contributions from the terms involving $\mu_1$ and $\mu_2$ in \C{calcmore} 
to obtain their dependence on the zeta functions. To do so, we proceed 
exactly along the lines of~\cite{Green:2005ba}, and so we mention only the results. 

Mutiplying the equation involving ${\tilde f}^{(0,0)}_2$ by $E_7 (\tau, \bar\tau)$, and integrating over the fundamental domain
of $SL(2,\mathbb{Z})$ gives
\be a_5 = \frac{32}{13\pi^4} \sum_{n=1}^\infty \frac{1}{n^4} \mu (n,3/2) \mu (n, 5/2),\ee
where we have used both \C{expSL2} and \C{altrep}, the Rankin--Selberg formula \C{RSf}, and 
\be \int_0^\infty dx x^6 K_1 (x) K_2 (x) = \frac{32}{5}.\ee  
Similarly, multiplying the equation involving ${\tilde f}^{(0,0)}_1$ by $E_5 (\tau, \bar\tau)$ and using
\be \int_0^\infty dx x^4 K_1 (x) K_2 (x) = 2,\ee
we find
\be a_4 = \frac{4}{27\pi^2} \sum_{n=1}^\infty \frac{1}{n^2} \mu (n,3/2) \mu (n, 5/2).\ee
Finally, we use Ramanujan's formula~\cite{Apostol}
\be \sum_{n=1}^\infty \frac{1}{n^r} \mu (n,s) \mu (n, s') = \frac{\zeta (r) \zeta (r + 2s - 1)
\zeta (r + 2s' -1) \zeta (r + 2s + 2s' -2)}{\zeta (2r + 2s + 2s' -2)}\ee
to obtain
\be \label{val4,5}
a_4 =\frac{11}{4050} \pi^2 \zeta (6), \quad a_5 = \frac{8}{2835} \pi^2 \zeta (8).\ee
Taking into account only the terms involving $\mu_1$ and $\mu_2$ in~\C{calcmore}, the 
non--perturbative contributions to ${\tilde f}^{(0,0)}_i$ can also be evaluated along the lines of~\cite{Green:2005ba}. These correspond to D-instantons effects as well as contributions from D-instanton/D-anti--instanton pairs.

As a consistency check for the genus five amplitude, consider the four graviton amplitude
in $d=11$ supergravity at one loop on $T^2$~\cite{Green:1999pu}. Using~\C{treeamp}\ to fix relative normalizations, 
we obtain terms in the amplitude in string frame given by
\be \alpha'^5 \Big(\frac{\zeta(3) \zeta (5)}{1280} \tau_2^2 (s^5 + t^5 + u^5) + \ldots +
\frac{2\pi^2}{5} \zeta (8) \tau_2^{-8} \mathcal{W}^5 (s,t,u) \Big) \mathcal{R}^4,\ee
where
\be \mathcal{W}^5 (s,t,u) = \mathcal{G}_{st}^5 + \mathcal{G}_{su}^5 + \mathcal{G}_{tu}^5 
= \frac{1}{216216} (s^5 + t^5 + u^5)\ee
after using 
\be \mathcal{G}_{st}^5 = \int_0^1 d \omega_3 \int_0^{\omega_3} d \omega_2 \int_0^{\omega_2} d \omega_1 
\Big( s \omega_1 (\omega_3 -\omega_2) + t (\omega_2 - \omega_1) (1-\omega_3) \Big)^5. \ee
Hence the genus five contribution in~\C{val4,5}, which is proportional to $\pi^2 \zeta (8)$, 
is consistent with considerations of supergravity and duality. Similarly, we expect the genus four amplitude to
be proportional to $\pi^2 \zeta (6)$.

This structure gets more involved for higher values of $k$ 
where both the $D^{2k} \mathcal{R}^4$ and the
$\hat{G}^{2k} \lambda^{16}$ interactions have different space-time structures. The first case occurs for $k=6$ where the 
$D^{12} \mathcal{R}^4$ interaction yields both 
$$(s^2 + t^2 + u^2)^3 \mathcal{R}^4 \quad {\rm and} \quad (s^3 + t^3 + u^3)^2 \mathcal{R}^4$$ 
when 
expressed in momentum space. Some or all of the different space-time structures 
can give different modular forms. However, for the $D^{2k} \mathcal{R}^4$ interaction, the modular form that multiplies a 
particular space-time structure can further split into a sum of independent 
modular forms as we saw above. This cannot happen for the $\hat{G}^{2k} \lambda^{16}$ interaction 
based on our general analysis. 

From considerations of the four graviton amplitude at one loop in $d=11$ supergravity, one can argue that the 
$D^{2k} \mathcal{R}^4$ interaction does not receive perturbative contributions beyond genus $k$ in type IIA string 
theory~\cite{Green:2006gt}. The same is true in type IIB string theory if we restrict to the $t_8 t_8 D^{2k} R^4$ part of the amplitude. 
So for $k \geq 5$, we see that the source terms in \C{karbone}
and \C{karbtwo} contribute up to genus $k-2$. The Laplace equations must then provide the genus $k-1$ and $k$
contributions. So there must be at least two independent modular forms of a given weight. 

Let us denote the  two modular forms which receive
contributions at genus $k-1$ (but not at genus $k$) and $k$ (but not at genus $k-1$) by $f^{(11+k, -11 -k)}_1$ and $f^{(11+k, -11 -k)}_2$, respectively. Then the behavior
\be f^{(11+k, -11 -k)}_1 \sim \tau_2^{(7-3k)/2}, \qquad f^{(11+k, -11 -k)}_2 \sim \tau_2^{3(1-k)/2}\ee
constrains~\C{karbone}\ and~\C{karbtwo} as follows
\bea \bar{D}_{-(12+k)} D_{11+k} f^{(11+k, -11 -k)}_1 = \frac{(k-29)(5k+17)}{16} f^{(11+k, -11 -k)}_1
+ {\rm sources}, \non \\
\bar{D}_{-(12+k)} D_{11+k} f^{(11+k, -11 -k)}_2 = \frac{(k-25)(5k+21)}{16} f^{(11+k, -11 -k)}_2
+ {\rm sources}.\eea
This leads us to conjecture that the genus $k-1$ coefficient of $D^{2k} {\mathcal R}^4$ is $\pi^2 \zeta (2k-4)$ up to a rational proportionality constant for $k \geq 3$. This is proportional to the genus $k-1$ contribution to $D^{2(k-1)} {\mathcal R}^4$ as shown in~\cite{Green:2006gt}.

\section*{Acknowledgements}
 
We would like to thank M.~B. Green, O.~Lunin and A.~Sinha for useful comments. A.~B. would also like to thank the Enrico Fermi Institute for hospitality during the final stage of this project. S.~S. would also like to thank the Aspen Center for Physics for hospitality during the completion of this project. 
The work of A.~B. is supported in part by NSF Grant No.~PHY-0503584 and the William D. Loughlin membership.
The work of S.~S. is supported in part by NSF CAREER Grant No. PHY-0758029 and by NSF Grant No. 0529954.

%%%%%%%%%%%%%%%%%%%%%%%%%%%%%%%%%%%%%%%%%%
%%%%%%%%%%%%%%%%%%%%%%%%%%%%%%%%%%%%%%%%%%

\newpage
\appendix

\section{Useful Formulae from Type IIB Supergravity and Some Spinorial Identities}
\label{susysummary}

The spinors in type IIB string theory are chiral spinors. The dilatino, $\lambda$, and the gravitino, $\psi_\mu$, have
opposite chiralities while the supersymmetry parameter, $\epsilon$, has the same chirality as the gravitino.

The relevant linearized supersymmetry transformations are
\bea &\delta^{(0)} \tau &= 2 \tau_2 \bar\epsilon^* \lambda, \qquad \delta^{(0)} \bar\tau
=- 2 \tau_2 \bar\epsilon \lambda^*,\non \\ & \delta^{(0)} e_\mu^m &= i (\bar\epsilon
\gamma^m \psi_\mu + \bar\epsilon^* \gamma^m \psi_\mu^*) , \non \\
& \delta^{(0)} \lambda &= i\gamma^\mu \epsilon^* \hat{P}_\mu  -\frac{i}{24} \gamma^{\mu\nu\rho} 
\epsilon \hat{G}_{\mu\nu\rho} + \frac{3}{4} i \lambda (\bar\epsilon \lambda^*) - \frac{3}{4} i
\lambda (\bar\epsilon^* \lambda), \non \\ & \delta^{(0)} \psi_\mu &= 
D_\mu \epsilon +\frac{i}{480} \g^{\rho_1 \ldots \rho_5} \g_\mu \epsilon
\hat{F}_{\rho_1 \ldots \rho_5}  + \frac{1}{96} \Big( \gamma_\mu^{~\nu\rho\lambda} \hat{G}_{\nu\rho\lambda}
-  9 \gamma^{\rho\lambda} \hat{G}_{\mu\rho\lambda} \Big) \epsilon^* \non \\ &&-\frac{7}{16} 
\Big( \gamma_\rho \lambda \bar\psi_\mu \gamma^\rho \epsilon^*
- \frac{1}{1680} \gamma_{\rho_1 \ldots \rho_5} \lambda \bar\psi_\mu \gamma^{\rho_1 \ldots \rho_5}
\epsilon^* \Big) \non \\ &&+\frac{i}{32} \Big[ \Big( \frac{9}{4} \gamma_\mu \gamma^\rho + 3 \g^\rho \g_\mu \Big)
\epsilon \bar\lambda \g_\rho \lambda \non \\ &&- \Big( \frac{1}{24} \g_\mu \g^{\rho_1\rho_2\rho_3} + \frac{1}{6}
\g^{\rho_1\rho_2\rho_3} \g_\mu \Big) \epsilon \bar\lambda
\g_{\rho_1\rho_2\rho_3} \lambda +\frac{1}{960} \g_\mu \g^{\rho_1 \ldots \rho_5} \epsilon
 \bar\lambda \g_{\rho_1 \ldots \rho_5} \lambda \Big] \non \\ &&+ \frac{1}{4} i\psi_\mu 
(\bar\epsilon \lambda^*) -\frac{1}{4} i \psi_\mu (\bar\epsilon^* \lambda),\eea
where 
\bea \label{defform}
&&\hat{P}_\mu = \frac{i\p_\mu \tau}{2\tau_2} - \bar\psi^*_\mu \lambda, \non \\
&&\hat{G}_{\mu\nu\rho} = G_{\mu\nu\rho} - 3 \bar\psi_{[\mu}
\gamma_{\nu\rho]} \lambda - 6i \bar\psi^*_{[\mu} \gamma_\nu 
\psi_{\rho ]}, \non \\
&& \hat{F}_{5~\mu_1 \ldots \mu_5} = F_{5~\mu_1 \ldots \mu_5} -5 \bar\psi_{[\mu_1} \g_{\mu_2 \mu_3 \mu_4} \psi_{\mu_5]}
- \frac{1}{16} \bar\lambda \g_{\mu_1 \ldots \mu_5} \lambda. \eea
There are useful relations for the dilatinos:
\bea & (\lambda^r )_{a_{r+1} \cdots a_{16}} &= \frac{1}{r!} \epsilon_{a_1 \cdots a_{16}} \lambda^{a_1}
\cdots \lambda^{a_r} ,\non \\
& (\lambda^{14})_{ab} \lambda_c \lambda_d &= \lambda^{16} (\delta_{ac} \delta _{bd} - \delta_{ad} \delta_{bc}),\non \\
&(\lambda^{14})_{ab} \lambda_c & = (\lambda^{15})_b \delta_{ac} - (\lambda^{15})_a \delta_{bc},\non \\
& (\lambda^{15})_a \lambda^b & = \delta_a^b \lambda^{16}.\eea
Our metric has signature mostly plusses and the gamma matrices are real with the transpose given by
\be \g^0 \g^\mu = -(\g^\mu )^T \g^0 .\ee
Some useful relations involving gamma matrices are summarized below:
\bea & {\rm Tr}  (\g^{\mu_1 \mu_2 \mu_3} \g_{\nu_1\nu_2 \nu_3}) = -16 \Big( \delta^{\mu_1}_{\nu_1} \delta^{\mu_2}_{\nu_2} \delta^{\mu_3}_{\nu_3}
-\delta^{\mu_2}_{\nu_1} \delta^{\mu_1}_{\nu_2} \delta^{\mu_3}_{\nu_3}
\non \\ 
& + \delta^{\mu_2}_{\nu_1} \delta^{\mu_3}_{\nu_2} 
\delta^{\mu_1}_{\nu_3} -\delta^{\mu_3}_{\nu_1} \delta^{\mu_2}_{\nu_2} \delta^{\mu_1}_{\nu_3} +\delta^{\mu_3}_{\nu_1} 
\delta^{\mu_1}_{\nu_2} \delta^{\mu_2}_{\nu_3}
-\delta^{\mu_1}_{\nu_1} \delta^{\mu_3}_{\nu_2} \delta^{\mu_2}_{\nu_3} \Big), &
\non \\ 
& \g^\mu \g_\nu \g_\mu = -8 \g_\nu, \qquad
\g^\mu \g_{\nu\rho\s} \g_\mu = -4 \g_{\nu\rho\s} , \qquad
\g_{\mu\nu}\g^{\mu\nu\rho} = -72 \g^\rho, &
 \\  
 & \g^\mu \g_{\nu_1 \cdots \nu_5} \g_\mu =0, \qquad
 \g^{\mu\nu\rho} \g_\s \g_{\mu\nu\rho} = 288 \g_\s,
\qquad \g^{\mu\nu\rho} \g_{\s_1 \s_2}  \g_{\mu\nu\rho} = -48 \g_{\s_1 \s_2}, &
\non \\
& \g^{\mu\nu\rho} \g_{\s_1 \s_2 \s_3} \g_{\mu\nu\rho} = -48 \g_{\s_1 \s_2 \s_3} ,\qquad
\g^{\mu\nu\rho} \g_{\s_1 \cdots\s_4} \g_{\mu\nu\rho} = 48 \g_{\s_1 \cdots \s_4} , \qquad
\g^{\mu\nu\rho} \g_{\s_1 \cdots \s_5} \g_{\mu\nu\rho} = 0, &
\non \\
& \g^{\mu_1 \cdots \mu_5} \g_{\s_1 \cdots \s_5} \g_{\mu_1 \cdots \mu_5} = 0,\qquad
\g^{\mu_1 \mu_2 \mu_3 \mu_4 \mu_5} \g_{\mu_2 \mu_3 \mu_4} = 336 \g^{\mu_1 \mu_5}, \qquad
\g^{\mu\nu} \g_{\mu\nu} = -90, &
\non \\ 
& \g^{\mu\nu\rho} \g_{\mu\nu\rho} = -720,\qquad
\g^{\mu_1 \mu_2 \mu_3 \mu_4} \g_{\mu_1 \mu_2 \mu_3 \mu_4} = 5040, \qquad
\g^{\mu_1 \mu_2 \mu_3 \mu_4 \mu_5} \g_{\mu_1 \mu_2 \mu_3 \mu_4 \mu_5} = 30240. & \non\eea
Conjugation for the spinors is defined  by
\be (\psi_a \chi_b)^* = -\psi_a^* \chi_b^* .\ee
For spinors $(\lambda_1, \lambda_2, \lambda_3)$ of the same chirality, we have the relations
\bea \label{vanspin}
\g^{\mu_1 \cdots \mu_5} \lambda_1 (\bar\lambda_2 \g_{\mu_1 \cdots \mu_5} \lambda_3) =0, \non \\
\bar\lambda_1 \g^\mu \lambda_2 = -\bar\lambda^*_2 \g^\mu \lambda^*_1 , \non \\
\bar\lambda_1 \g^{\mu\nu\rho} \lambda_2 = \bar\lambda^*_2 \g^{\mu\nu\rho} \lambda^*_1. \eea

We extensively use the Fierz identity involving two spinors $\lambda_1$ and $\lambda_2$ of the same
chirality
\be \lambda_1^a \bar\lambda_2^b = -\frac{1}{16} \bar\lambda_2 \g^\mu \lambda_1 \g_\mu^{ab} 
+\frac{1}{96} \bar\lambda_2 \g^{\mu\nu\rho} \lambda_1 \g_{\mu\nu\rho}^{ab} 
-\frac{1}{3840} \bar\lambda_2 \g^{\mu_1 \cdots \mu_5} \lambda_1 \g_{\mu_1 \cdots \mu_5}^{ab} .\ee

\newpage

\section{Useful Properties of Modular Forms under $SL(2,\Z)$}
\label{summarymodular}

A modular form $\Phi^{(m,n)} (\tau, \bar\tau)$ of weight $(m,n)$ transforms under the $SL(2,\mathbb{Z})$ transformation
\be \tau \rightarrow \frac{a\tau +b}{c\tau +d},\ee
according to the rule
\be \label{modformvar}
\Phi^{(m,n)} (\tau, \bar\tau) \rightarrow (c\tau +d)^m (c\bar\tau +d)^n \Phi^{(m,n)} (\tau, \bar\tau).\ee 
We define modular covariant derivatives as
\bea \label{modcovder}
D_m = i \Big( \tau_2 \frac{\p}{\p \tau} - i\frac{m}{2} \Big), \non \\ \bar{D}_n = 
-i \Big( \tau_2 \frac{\p}{\p \bar\tau} + i \frac{n}{2} \Big),\eea
which take
\be D_m \Phi^{(m,n)} \rightarrow \Phi^{(m+1, n-1)}, \qquad \bar{D}_n \Phi^{(m,n)}
\rightarrow \Phi^{(m-1, n+1)}.\ee

The modular invariant non--holomorphic Eisenstein series of order $s$ for $SL(2,\mathbb{Z})$ 
is defined by~\cite{AT}
\bea \label{expSL2}
E_{s} (\tau,\bar{\tau})  &=& \sum_{(p,q) \neq (0,0)} 
\frac{\tau_2^{s}}{\vert p + q \tau\vert^{2s }} \non \\
&=& 2 \zeta(2s) \tau_2^s + 2 \sqrt{\pi} \tau_2^{1-s} \frac{\Gamma (s -1/2)}{\Gamma (s)}
\zeta (2s -1) \non \\ &&+ \frac{4 \pi^s \sqrt{\tau_2}}{\Gamma (s)} \sum_{k \neq 0} 
\vert k \vert^{s - 1/2} \mu (k,s) K_{s - 1/2} (2\pi \tau_2 \vert k
\vert) e^{2\pi i k \tau_1}, \eea
where
\be \mu (k,s) = \sum_{m > 0, m | k} \frac{1}{m^{2s -1}}.\ee
We also use the representation
\be \label{altrep}
E_{s} (\tau,\bar{\tau}) = 2 \zeta (2s) \sum_{\gamma \in \Gamma_\infty  \setminus SL(2,\mathbb{Z})} {\rm Im}
(\gamma \cdot \tau)^s ,\ee
and the Rankin--Selberg formula~\cite{AT}
\be \label{RSf}
\int_{\mathcal{F}} \frac{d^2 \tau}{\tau_2^2} \sum_{\gamma \in \Gamma_\infty  \setminus SL(2,\mathbb{Z})} \psi (\gamma 
\cdot \tau) f (\tau) = \int_0^\infty \frac{d \tau_2}{\tau_2^2} \psi (\tau_2) \int_{-1/2}^{1/2} d \tau_1 f (\tau), \ee
where $\mathcal{F}$ is the fundamental domain of $SL(2,\mathbb{Z})$.

%%%%%%%%%%%%%%%%%%%%%%%%%%%%%%%%%%%%%%%%%%
%%%%%%%%%%%%%%%%%%%%%%%%%%%%%%%%%%%%%%%%%%

%%%%%%%%%%%%%%%%%%%%%%%%%%%%%%%%%%%%%%%%%%
%%%%%%%%%%%%%%%%%%%%%%%%%%%%%%%%%%%%%%%%%%

%\newpage
%%%%%%%%%%%%%%%%%%%%%%%%%%%%%%%%%%%%%%%%%%%%%%%%%%%%%%%%%%%%

%\bibliographystyle{utphys}
%\bibliography{myrefs}

\begin{thebibliography}{10}

\bibitem{Berkovits:2006vc}
N.~Berkovits, ``{New higher-derivative $R^4$ theorems},'' {\em Phys. Rev.
  Lett.} {\bf 98} (2007) 211601,
\href{http://www.arXiv.org/abs/hep-th/0609006}{{\tt hep-th/0609006}}.
%%CITATION = HEP-TH/0609006;%%.

\bibitem{Green:1997tv}
M.~B. Green and M.~Gutperle, ``{Effects of D-instantons},'' {\em Nucl. Phys.}
  {\bf B498} (1997) 195--227,
\href{http://www.arXiv.org/abs/hep-th/9701093}{{\tt hep-th/9701093}}.
%%CITATION = HEP-TH/9701093;%%.

\bibitem{Green:1997di}
M.~B. Green and P.~Vanhove, ``{D-instantons, strings and M-theory},'' {\em
  Phys. Lett.} {\bf B408} (1997) 122--134,
\href{http://www.arXiv.org/abs/hep-th/9704145}{{\tt hep-th/9704145}}.
%%CITATION = HEP-TH/9704145;%%.

\bibitem{Green:1997as}
M.~B. Green, M.~Gutperle, and P.~Vanhove, ``{One loop in eleven dimensions},''
  {\em Phys. Lett.} {\bf B409} (1997) 177--184,
\href{http://www.arXiv.org/abs/hep-th/9706175}{{\tt hep-th/9706175}}.
%%CITATION = HEP-TH/9706175;%%.

\bibitem{Kiritsis:1997em}
E.~Kiritsis and B.~Pioline, ``{On $R^4$ threshold corrections in type IIB
  string theory and (p,q) string instantons},'' {\em Nucl. Phys.} {\bf B508}
  (1997) 509--534,
\href{http://www.arXiv.org/abs/hep-th/9707018}{{\tt hep-th/9707018}}.
%%CITATION = HEP-TH/9707018;%%.

\bibitem{Green:1997me}
M.~B. Green, M.~Gutperle, and H.-h. Kwon, ``{$\lambda^{16}$ and related terms
  in M-theory on $T^2$},'' {\em Phys. Lett.} {\bf B421} (1998) 149--161,
\href{http://www.arXiv.org/abs/hep-th/9710151}{{\tt hep-th/9710151}}.
%%CITATION = HEP-TH/9710151;%%.

\bibitem{Pioline:1998mn}
B.~Pioline, ``{A note on non-perturbative $R^4$ couplings},'' {\em Phys. Lett.}
  {\bf B431} (1998) 73--76,
\href{http://www.arXiv.org/abs/hep-th/9804023}{{\tt hep-th/9804023}}.
%%CITATION = HEP-TH/9804023;%%.

\bibitem{Green:1998by}
M.~B. Green and S.~Sethi, ``{Supersymmetry constraints on type IIB
  supergravity},'' {\em Phys. Rev.} {\bf D59} (1999) 046006,
\href{http://www.arXiv.org/abs/hep-th/9808061}{{\tt hep-th/9808061}}.
%%CITATION = HEP-TH/9808061;%%.

\bibitem{Green:1999pu}
M.~B. Green, H.-h. Kwon, and P.~Vanhove, ``{Two loops in eleven dimensions},''
  {\em Phys. Rev.} {\bf D61} (2000) 104010,
\href{http://www.arXiv.org/abs/hep-th/9910055}{{\tt hep-th/9910055}}.
%%CITATION = HEP-TH/9910055;%%.

\bibitem{Obers:1999es}
N.~A. Obers and B.~Pioline, ``{Eisenstein series in string theory},'' {\em
  Class. Quant. Grav.} {\bf 17} (2000) 1215--1224,
\href{http://www.arXiv.org/abs/hep-th/9910115}{{\tt hep-th/9910115}}.
%%CITATION = HEP-TH/9910115;%%.

\bibitem{Sinha:2002zr}
A.~Sinha, ``{The $\hat{G}^4 \lambda^{16}$ term in IIB supergravity},'' {\em
  JHEP} {\bf 08} (2002) 017,
\href{http://www.arXiv.org/abs/hep-th/0207070}{{\tt hep-th/0207070}}.
%%CITATION = HEP-TH/0207070;%%.

\bibitem{Berkovits:2004px}
N.~Berkovits, ``{Multiloop amplitudes and vanishing theorems using the pure
  spinor formalism for the superstring},'' {\em JHEP} {\bf 09} (2004) 047,
\href{http://www.arXiv.org/abs/hep-th/0406055}{{\tt hep-th/0406055}}.
%%CITATION = HEP-TH/0406055;%%.

\bibitem{DHoker:2005jc}
E.~D'Hoker and D.~H. Phong, ``{Two-Loop Superstrings VI: Non-Renormalization
  Theorems and the 4-Point Function},'' {\em Nucl. Phys.} {\bf B715} (2005)
  3--90,
\href{http://www.arXiv.org/abs/hep-th/0501197}{{\tt hep-th/0501197}}.
%%CITATION = HEP-TH/0501197;%%.

\bibitem{DHoker:2005ht}
E.~D'Hoker, M.~Gutperle, and D.~H. Phong, ``{Two-loop superstrings and
  S-duality},'' {\em Nucl. Phys.} {\bf B722} (2005) 81--118,
\href{http://www.arXiv.org/abs/hep-th/0503180}{{\tt hep-th/0503180}}.
%%CITATION = HEP-TH/0503180;%%.

\bibitem{Green:2005ba}
M.~B. Green and P.~Vanhove, ``{Duality and higher derivative terms in M
  theory},'' {\em JHEP} {\bf 01} (2006) 093,
\href{http://www.arXiv.org/abs/hep-th/0510027}{{\tt hep-th/0510027}}.
%%CITATION = HEP-TH/0510027;%%.

\bibitem{Green:2006gt}
M.~B. Green, J.~G. Russo, and P.~Vanhove, ``{Non-renormalisation conditions in
  type II string theory and maximal supergravity},'' {\em JHEP} {\bf 02} (2007)
  099,
\href{http://www.arXiv.org/abs/hep-th/0610299}{{\tt hep-th/0610299}}.
%%CITATION = HEP-TH/0610299;%%.

\bibitem{Basu:2007ru}
A.~Basu, ``{The $D^4 R^4$ term in type IIB string theory on $T^2$ and U-
  duality},''
\href{http://www.arXiv.org/abs/arXiv:0708.2950 [hep-th]}{{\tt arXiv:0708.2950
  [hep-th]}}.
%%CITATION = ARXIV:0708.2950;%%.

\bibitem{Basu:2007ck}
A.~Basu, ``{The $D^6 R^4$ term in type IIB string theory on $T^2$ and U-
  duality},''
\href{http://www.arXiv.org/abs/0712.1252}{{\tt 0712.1252}}.
%%CITATION = 0712.1252;%%.

\bibitem{Green:1999qt}
M.~B. Green, ``{Interconnections between type II superstrings, M theory and $N
  = 4$ Yang-Mills},''
\href{http://www.arXiv.org/abs/hep-th/9903124}{{\tt hep-th/9903124}}.
%%CITATION = HEP-TH/9903124;%%.

\bibitem{Schwarz:1983qr}
J.~H. Schwarz, ``{Covariant Field Equations of Chiral N=2 D=10 Supergravity},''
  {\em Nucl. Phys.} {\bf B226} (1983)
269.
%%CITATION = NUPHA,B226,269;%%.

\bibitem{Howe:1983sra}
P.~S. Howe and P.~C. West, ``{The Complete N=2, D=10 Supergravity},'' {\em
  Nucl. Phys.} {\bf B238} (1984)
181.
%%CITATION = NUPHA,B238,181;%%.

\bibitem{Paban:1998ea}
S.~Paban, S.~Sethi, and M.~Stern, ``{Constraints from extended supersymmetry in
  quantum mechanics},'' {\em Nucl. Phys.} {\bf B534} (1998) 137--154,
\href{http://www.arXiv.org/abs/hep-th/9805018}{{\tt hep-th/9805018}}.
%%CITATION = HEP-TH/9805018;%%.

\bibitem{Paban:1998qy}
S.~Paban, S.~Sethi, and M.~Stern, ``{Supersymmetry and higher derivative terms
  in the effective action of Yang-Mills theories},'' {\em JHEP} {\bf 06} (1998)
  012,
\href{http://www.arXiv.org/abs/hep-th/9806028}{{\tt hep-th/9806028}}.
%%CITATION = HEP-TH/9806028;%%.

\bibitem{Sethi:1999qv}
S.~Sethi and M.~Stern, ``{Supersymmetry and the Yang-Mills effective action at
  finite N},'' {\em JHEP} {\bf 06} (1999) 004,
\href{http://www.arXiv.org/abs/hep-th/9903049}{{\tt hep-th/9903049}}.
%%CITATION = HEP-TH/9903049;%%.

\bibitem{Sethi:2004np}
S.~Sethi, ``{Structure in supersymmetric Yang-Mills theory},'' {\em JHEP} {\bf
  10} (2004) 001,
\href{http://www.arXiv.org/abs/hep-th/0404056}{{\tt hep-th/0404056}}.
%%CITATION = HEP-TH/0404056;%%.

\bibitem{Berkovits:1994vy}
N.~Berkovits and C.~Vafa, ``{N=4 topological strings},'' {\em Nucl. Phys.} {\bf
  B433} (1995) 123--180,
\href{http://www.arXiv.org/abs/hep-th/9407190}{{\tt hep-th/9407190}}.
%%CITATION = HEP-TH/9407190;%%.

\bibitem{Ooguri:1995cp}
H.~Ooguri and C.~Vafa, ``{All loop N=2 string amplitudes},'' {\em Nucl. Phys.}
  {\bf B451} (1995) 121--161,
\href{http://www.arXiv.org/abs/hep-th/9505183}{{\tt hep-th/9505183}}.
%%CITATION = HEP-TH/9505183;%%.

\bibitem{Berkovits:1998ex}
N.~Berkovits and C.~Vafa, ``{Type IIB $R^4 H^{(4g-4)}$ conjectures},'' {\em
  Nucl. Phys.} {\bf B533} (1998) 181--198,
\href{http://www.arXiv.org/abs/hep-th/9803145}{{\tt hep-th/9803145}}.
%%CITATION = HEP-TH/9803145;%%.

\bibitem{GSW1}
M.~B. Green, J.~H. Schwarz, and E.~Witten, ``{Superstring Theory: Volume 1,
  Introduction},'' {\em Cambridge University Press} (1987).

\bibitem{VanNieuwenhuizen:1981ae}
P.~Van~Nieuwenhuizen, ``{Supergravity},'' {\em Phys. Rept.} {\bf 68} (1981)
189--398.
%%CITATION = PRPLC,68,189;%%.

\bibitem{Green:1981xx}
M.~B. Green and J.~H. Schwarz, ``{Supersymmetrical Dual String Theory. 2.
  Vertices and Trees},'' {\em Nucl. Phys.} {\bf B198} (1982)
252--268.
%%CITATION = NUPHA,B198,252;%%.

\bibitem{DHoker:1988ta}
E.~D'Hoker and D.~H. Phong, ``The geometry of string perturbation theory,''
  {\em Rev. Mod. Phys.} {\bf 60} (1988)
917.
%%CITATION = RMPHA,60,917;%%.

\bibitem{Green:1999pv}
M.~B. Green and P.~Vanhove, ``{The low energy expansion of the one-loop type II
  superstring amplitude},'' {\em Phys. Rev.} {\bf D61} (2000) 104011,
\href{http://www.arXiv.org/abs/hep-th/9910056}{{\tt hep-th/9910056}}.
%%CITATION = HEP-TH/9910056;%%.

\bibitem{Green:1982sw}
M.~B. Green, J.~H. Schwarz, and L.~Brink, ``{N=4 Yang-Mills and N=8
  Supergravity as Limits of String Theories},'' {\em Nucl. Phys.} {\bf B198}
  (1982)
474--492.
%%CITATION = NUPHA,B198,474;%%.

\bibitem{Russo:1997mk}
J.~G. Russo and A.~A. Tseytlin, ``{One-loop four-graviton amplitude in
  eleven-dimensional supergravity},'' {\em Nucl. Phys.} {\bf B508} (1997)
  245--259,
\href{http://www.arXiv.org/abs/hep-th/9707134}{{\tt hep-th/9707134}}.
%%CITATION = HEP-TH/9707134;%%.

\bibitem{Bershadsky:1993cx}
M.~Bershadsky, S.~Cecotti, H.~Ooguri, and C.~Vafa, ``{Kodaira-Spencer theory of
  gravity and exact results for quantum string amplitudes},'' {\em Commun.
  Math. Phys.} {\bf 165} (1994) 311--428,
\href{http://www.arXiv.org/abs/hep-th/9309140}{{\tt hep-th/9309140}}.
%%CITATION = HEP-TH/9309140;%%.

\bibitem{Basu:2006cs}
A.~Basu, ``{The $D^10$ $R^4$ term in type IIB string theory},'' {\em Phys.
  Lett.} {\bf B648} (2007) 378--382,
\href{http://www.arXiv.org/abs/hep-th/0610335}{{\tt hep-th/0610335}}.
%%CITATION = HEP-TH/0610335;%%.

\bibitem{Chalmers:2006aj}
G.~Chalmers, ``{Integrability in string theories},''
\href{http://www.arXiv.org/abs/physics/0604016}{{\tt physics/0604016}}.
%%CITATION = PHYSICS/0604016;%%.

\bibitem{Green:2008bf}
M.~B. Green, J.~G. Russo, and P.~Vanhove, ``{Modular properties of two-loop
  maximal supergravity and connections with string theory},''
\href{http://www.arXiv.org/abs/0807.0389}{{\tt 0807.0389}}.
%%CITATION = 0807.0389;%%.

\bibitem{Green:2008uj}
M.~B. Green, J.~G. Russo, and P.~Vanhove, ``{Low energy expansion of the
  four-particle genus-one amplitude in type II superstring theory},'' {\em
  JHEP} {\bf 02} (2008) 020,
\href{http://www.arXiv.org/abs/0801.0322}{{\tt 0801.0322}}.
%%CITATION = 0801.0322;%%.

\bibitem{Berkovits:2006vi}
N.~Berkovits and N.~Nekrasov, ``{Multiloop superstring amplitudes from
  non-minimal pure spinor formalism},'' {\em JHEP} {\bf 12} (2006) 029,
\href{http://www.arXiv.org/abs/hep-th/0609012}{{\tt hep-th/0609012}}.
%%CITATION = HEP-TH/0609012;%%.

\bibitem{Apostol}
T.~M. Apostol, ``Introduction to analytic number theory,'' {\em Springer
  Verlag, New York} (1976).

\bibitem{AT}
A.~Terras, ``{Harmonic Analysis on Symmetric Spaces and Applications, Vols. I,
  II},'' {\em Springer-Verlag, N.Y.} (1988).

\end{thebibliography}
%\providecommand{\href}[2]{#2}\begingroup\raggedright\begin{thebibliography}{10}

\providecommand{\href}[2]{#2}\begingroup\raggedright\endgroup

\end{document}